\documentclass{myaa}
\usepackage {graphicx}
\usepackage {amssymb}
\usepackage {mathrsfs}
\usepackage {txfonts}
\usepackage {balance}
\usepackage{afterpage}
\def\Teff  {T$_{\mbox {\scriptsize eff}}$}
\def\teff  {T$_{\mbox {\scriptsize eff}}$}
\def\logg  {$\log g$}
\def\vt  {$v_t$}
\def\kms   {$ {\rm km \: s^{-1}  } $}
\def\12c13 {$ {\rm ^{12}C/^{13}C } $}
\def\12c   {$ {\rm ^{12}C  } $}
\def\13c   {$ {\rm ^{13}C  } $}
\newcommand{\cobold}{\ensuremath{\mathrm{CO}^5\mathrm{BOLD}}}
\newcommand{\linfor}{Linfor3D}
\newcommand{\nod}{--}

\newcommand{\mlp}{\ensuremath{\alpha_{\mathrm{MLT}}}}

\idline{1}{1}

\begin {document}

%\thesaurus{06(08.01.1; 08.06.2; 08.05.3; 08.06.3; 08.16.3;  08.19.5)}
%\thesaurus{06
%  (08.01.1;    %abond
%   08.06.2;    % formation
%   08.05.3;    % evol
%   08.06.3;    %fund  param
%   08.16.3     % Pop II
%   08.19.5)}   %supernovae

\title {First stars XII. Abundances in extremely metal-poor turnoff stars, 
and comparison with the giants. 
\thanks{Based on observations obtained with the ESO Very Large
Telescope at Paranal Observatory, Chile (Large Programme ``First
Stars'', ID 165.N-0276; P.I.: R. Cayrel, and Programme
078.B-0238; P.I.: M. Spite).}
}

\author {
P. Bonifacio\inst{1,2,3}\and
M. Spite\inst{2}\and 
R. Cayrel\inst{2}\and
V. Hill\inst{2,4}\and
F. Spite\inst{2}\and
P. Fran\c cois\inst{2}\and
B. Plez\inst{5,6}\and
H.-G Ludwig\inst{1,2}\and
E. Caffau\inst{2}\and
P.~Molaro\inst{2,3}\and
E. Depagne\inst{7}\and 
J. Andersen\inst{8,9}\and
B. Barbuy\inst{10}\and
T.C. Beers\inst{11}\and
B. Nordstr\"{o}m\inst{8}\and
F. Primas\inst{12}
}

\institute {
CIFIST Marie Curie Excellence Team
\and
GEPI, Observatoire de Paris, CNRS, Universit\'e Paris Diderot; Place
Jules Janssen 92190
Meudon, France \\
\email{[Piercarlo.Bonifacio;Monique.Spite;Roger.Cayrel;Vanessa.Hill;Francois.Spite;Patrick.Francois;\\
Hans.Ludwig;Elisabetta.Caffau]@obspm.fr}
         \and
Istituto Nazionale di Astrofisica - Osservatorio Astronomico di Trieste,
    Via Tiepolo 11, I-34143 Trieste, Italy\\
\email {molaro@oats.inaf.it}
\and
Laboratoire Cassiop{\'e}e UMR 6202, Universit{\'e} de 
Nice Sophia-Antipolis, CNRS, Observatoire de la C\^ote d’Azur
\email {Vanessa.Hilla@oca.eu}
\and 
   GRAAL, Universit\'e de Montpellier II, F-34095 Montpellier Cedex 
05, France\\
   \email {Bertrand.Plez@graal.univ-montp2.fr}
               \and
Department of Physics and Astronomy, Uppsala Astronomical Observatory, Box
515, S-751 20 Uppsala, Sweden
               \and
Las Cumbres Observatory, Goleta, CA 93117, USA\\
\email{edepagne@lcogt.net}
         \and
         The Niels Bohr Institute, Astronomy Group, Juliane Maries Vej 30,
         DK-2100 Copenhagen, Denmark\\
   \email {[ja;birgitta]@astro.ku.dk}
\and
     Nordic Optical Telescope, Apartado 474, E-38700 Santa Cruz de 
     La Palma, Spain\\
   \email {ja@not.iac.es}
\and
    IAG, Universidade de Sao Paulo, Depto. de Astronomia, 
    Rua do Matao 1226, Sao Paulo 05508-900, Brazil, 
   \email {barbuy@astro.iag.usp.br}
\and
Department of Physics \& Astronomy, CSCE: Center for the Study of
Cosmic
Evolution, and JINA: Joint Institute for Nuclear
Astrophysics, Michigan State University, East Lansing, MI 48824, USA
\email{beers@pa.msu.edu}
\and
  European Southern Observatory, Karl Schwarzschild-Str. 2, 
  D-85749 Garching bei M\"unchen, Germany
   \email {fprimas@eso.org}
}

\date {Received 16 July 2008/ Accepted 15 March 2009}
\titlerunning{First Stars XII. Detailed abundances in EMP dwarfs}
\authorrunning{P. Bonifacio et al.}

%%%% ABSTRACT %%%
\abstract
{The detailed chemical abundances of extremely metal-poor (EMP) stars are 
key guides to understanding the early chemical evolution of the Galaxy. 
Most existing data are, however, for giant stars which may have experienced 
internal mixing later. 
}
% aims heading (mandatory)
{We aim to compare the results for giants with new, accurate abundances for 
all observable elements in 18 EMP turnoff stars.
}
% methods heading (mandatory)
{VLT/UVES spectra at $R\sim45,000$ and S/N$\sim$ 130 per pixel
($\lambda\lambda$ 330-1000 nm) are analysed with OSMARCS model
atmospheres and the TURBOSPECTRUM code to derive abundances for C, Mg,
Si, Ca, Sc, Ti, Cr, Mn, Co, Ni, Zn, Sr, and Ba.
}
% results heading (mandatory)
{For Ca, Ni, Sr, and Ba, we find excellent consistency with our
earlier sample of EMP giants, at all metallicities.  However, our
abundances of C, Sc, Ti, Cr, Mn and Co are $\sim$0.2 dex larger than
in giants of similar metallicity.  Mg and Si abundances are $\sim$0.2
dex lower ( the giant [Mg/Fe] values are slightly revised), while Zn is 
again $\sim$0.4 dex higher than in giants of similar [Fe/H] (6 stars only).
}
% conclusions heading (optional)
{For C, the dwarf/giant discrepancy could possibly have an
astrophysical cause, but for the other elements it must arise from
shortcomings in the analysis.  Approximate computations of granulation
(3D) effects yield smaller corrections for giants than for dwarfs,
but suggest that this is an unlikely explanation, except perhaps for
C, Cr, and Mn.  NLTE computations for Na and Al provide consistent
abundances between dwarfs and giants, unlike the LTE results, and
would be highly desirable for the other discrepant elements as well.
Meanwhile, we recommend using the giant abundances as reference data
for Galactic chemical evolution models.
}
\keywords {Galaxy: abundances -- Galaxy: halo -- Galaxy: evolution -- 
Stars: abundances -- Stars: Population II -- Stars: Supernovae}
\maketitle
%
%________________________________________________________________
%%%% 1-INTRODUCTION %%%

%\balance
\section {Introduction} 

The surface composition of a cool star is a good diagnostic of the
chemical composition of the gas from which it formed, if mixing with
material processed inside the star itself has not occurred.  Cool,
long-lived stars have thus been extensively used to study the
early chemical evolution of our Galaxy (and, by implication, other
galaxies as well).
 The trends in abundance ratios which have been established over
the last 30 years provide important constraints on the early chemical 
evolution of the Milky Way (see Cayrel \cite{Cayrel1996}, 
\cite{Cayrel2006} for classic and recent reviews of the topic).

Our own programme, "First Stars", is a comprehensive spectroscopic
study of extremely metal-poor (EMP) stars to obtain precise
information on the chemical composition of the early ISM and the
yields of the first generation(s) of supernovae, conducted with the
VLT and UVES spectrograph.
The target stars have been selected from the medium-resolution
follow-up (Beers et al. in preparation; Allende Prieto
et al. \cite{allende}) of the HK objective-prism survey (Beers et al.
\cite{beers85,beers92} and Beers
\cite{beers99}), initiated by George Preston and Steve Shectman, and
later substantially extended and followed up by Beers as part of many
collaborations, including the present one.

Several papers have presented our results on the giant stars, which
lend themselves most readily to the study of many elements: Hill et
al.  (\cite {HPC02} - First Stars I), Depagne et al.  (\cite {DHS02} -
First Stars II), Fran\c cois et al.  (\cite {FDH03} - First Stars
III), Cayrel et al.  (\cite {CDS04} - First Stars V), Spite et al.
(\cite {SCP05} - First Stars VI), Fran\c cois et al.  (\cite{FDH06} -
First Stars VIII), and Spite et al.  (\cite {SCH06} - First Stars IX).
In these papers, we found the abundances in some giants to have been
altered with respect to their initial chemical composition, due to
mixing with layers affected by nuclear burning.  All the stars have
undergone the first dredge-up, so their abundances of Li, C, and N are
under suspicion.  However, our detailed analysis (First Stars VI and
IX), showed that the surface abundances of the less luminous giants (those below
the ``bump'' in the luminosity function) are not significantly affected
by mixing.

It is therefore expected that the less-luminous giants and dwarfs 
should display 
the same abundances, provided that the surface composition of the latter 
has not been changed by atmospheric phenomena, such as diffusion.
Comparing abundance ratios in dwarfs and giants can therefore, in 
principle, yield insight into the degree of mixing in giants and diffusion
in dwarfs as well as which element ratios are reliable guides to the 
composition of the early ISM in the Galaxy.

So far, only few of our papers have discussed results for EMP dwarfs: 
Sivarani et al.  (\cite {SBM04} - First Stars IV, \cite{SBB06} - First 
Stars X), Bonifacio et al.  (\cite {BMS06} - First Stars VII), and 
Gonz{\'a}lez Hern{\'a}ndez et al. (\cite{jonay} - First Stars XI).
First Stars VII focused on the Li abundance, but also discussed the 
model parameters and [Fe/H] of the dwarf sample in considerable detail. 
Here we discuss the abundances from C to Ba in the same stars and compare 
the results for dwarfs and giants.

\section {Observations and reduction}

\begin {figure}
\begin {center}
\resizebox{\hsize}{!}{\includegraphics[clip=true]{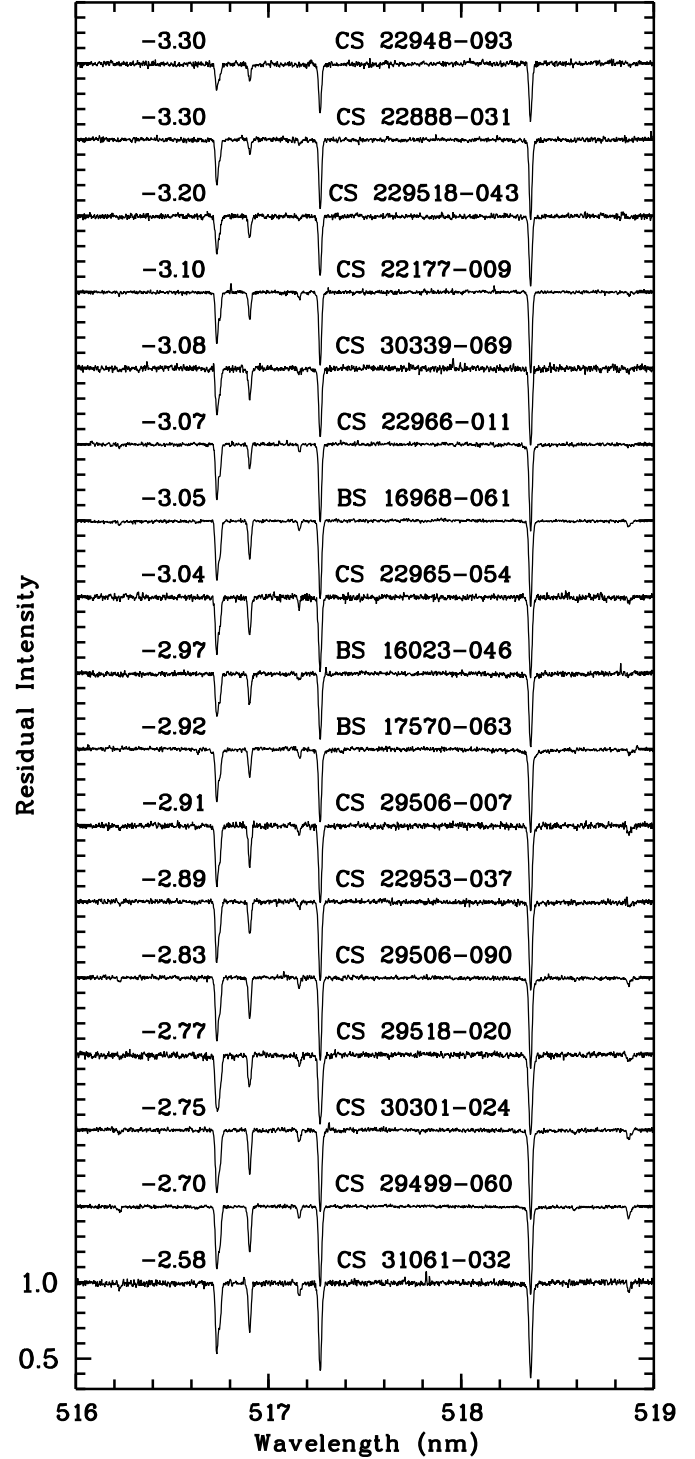}}
\caption{The region of the \ion{Mg}{i} b triplet in the program stars. 
[Fe/H] is shown to the left of each spectrum. In these EMP stars, the 
triplet lines have no damping wings.}
\label {Mgb}
\end {center}
\end {figure}

The sample of stars and the observational data are the same as discussed 
in Paper VII (Bonifacio et al.  \cite {BMS06}). 
The observations were performed with the ESO VLT and the
high-resolution spectrograph UVES (Dekker et al.  \cite {DDK00}) at a 
resolution of $R$= 45,000 and typical S/N ratios per pixel of 
$\sim$130  
on the coadded spectra
(average 5 pixels per resolution element). 
The spectra were 
reduced using the UVES context within MIDAS (Ballester et al.  
\cite{BMB00}); see paper V for details. The region of the \ion{Mg}{i} b 
triplet in our spectra is shown in Fig. \ref{Mgb} (see also Fig. 1 of 
Paper VII, which shows the Li line in the same stars).
Equivalent widths were measured on the coadded spectra. 
For a few stars, for which spectra with
     different resolutions (different slit-width, or image slicer used)
     were available, we coadded separately the spectra with the same
     resolution and then averaged the equivalent widths.

\section{Determination of atmospheric parameters} \label{analysis}

We have carried out a classical 1D LTE analysis using OSMARCS models
(see, e.g., Gustafsson et al.  \cite{GBE75}, \cite{GEE03}, \cite{G2008}).  
Estimates of \Teff\
were derived from the wings of H$\alpha$; log g estimates were obtained
by consideration of the ionisation equilibrium of iron. 
Microturbulent velocities were fixed by requiring no trend of 
[\ion{Fe}{i}/H] with equivalent width. Details are given in 
``First Stars VII", together with an extensive discussion of the 
effective temperature scale. 
In that paper we established that our H$\alpha$ based temperatures
satisfy the iron excitation equilibrium and are also in good agreement
with the Alonso et al. (\cite{alonso}) colour-temperature calibration,
which we used for the giant stars (Cayrel et al. \cite{CDS04}) 
The adopted parameters are listed in Table \ref {tabmod}. 

The parameters of the subgiant star BS\,16076-006 require a comment,  
because the Balmer line broadening in this star increases from H$\alpha$ 
towards the higher members of the Balmer series. Our adopted \teff\ 
(5199\,K) is derived from the wings of H$\alpha$, but the wings of H$\delta$ 
correspond to a much higher effective temperature, of the order of
5900\,K.
All values of \teff\ derived from colours are also consistently higher 
than derived from the H$\alpha$ profile, confirming this peculiarity.
This star was also analysed from medium-resolution ESI - Keck spectra 
($R$=7000) by Lai et al. (\cite{lai2004}), who adopted a \teff  = 5458\,K, 
based on photometry.
Such a \teff\ is compatible with the profile of H$\gamma$, but too low 
to reproduce the profile of H$\delta$. The reason for this peculiar behaviour
(e.g. a binary companion or chromospheric activity), needs further 
investigation, but the three radial velocities of derived from our two 
spectra and that of Lai et al. (\cite{lai2004}) show no evidence of variation.
None of our results depends critically on the abundances of this star,
however.

\section{Abundance determination}

The abundance analysis was performed using the LTE spectral line
analysis code "Turbospectrum" (Alvarez and Plez, \cite {AP98}).  The
abundances of the different elements have been determined mainly from
the equivalent widths of unblended lines.  However, synthetic spectra
have been used to determine abundances from the molecular bands, or in
cases when the lines were severely blended, affected by hyperfine
structure, or were strong enough to show significant damping wings
(see Sect. \ref{mag}).  The abundances of C and the $\alpha$ elements 
(as well as for Sc) are listed in Table \ref{tababund}, those of the 
heavier (neutron-capture) elements are listed
in Table \ref{tababund1}.

Abundance uncertainties are discussed in detail in Cayrel et al.
(\cite {CDS04}) and Bonifacio et al.  (\cite{BMS06}).  For a given
temperature, the ionisation equilibrium provides an estimate of the
gravity with an internal precision of about 0.1 dex in log $g$, and
the microturbulent velocity can be constrained within about 0.2 \kms.
The largest uncertainty comes from the temperature determination,
which is uncertain by $\sim$100K.

The total error estimate is not the quadratic sum of the various
sources of uncertainty, because the covariance terms are important.
As an illustration of the total expected uncertainty we have computed
the abundances of CS~29177-009 with different models: Model A has the
nominal temperature 6260~K, gravity (log $g$ = 4.5), and
microturbulent velocity ($v_t$ = 1.3 \kms), while Models B and C
differ in log $g$ and $v_t$ by 1$\sigma$.  Model D has a temperature
100~K lower and the same log $g$ and $v_t$, while in Model E we have
determined the ``best'' values of log $g$ and $v_t$ corresponding to
the lower temperature.  The detailed results of these computations are
given in Table \ref{errors}.

%FIGURE 1
\afterpage{\clearpage}

%TABLE 1
\begin {table}
\caption {Adopted model atmosphere parameters. Our UVES spectra show that
BS~16076-006 is in fact a subgiant (First Stars VII), and CS~29527-015 a 
double-lined binary; these stars are omitted in Fig. \ref{Mgb}. All the
others seem to be single turnoff stars. 
}
\label {tabmod}
\begin {center}
\begin{tabular}{llc@{ }c@{ }c@{ } c@{ }c@{ }c@{ }c@{ }c@{ }r@{ }r@{
}c@{ }c@{ }c@{ }c}
\hline 
\hline
&Star          &   $T_{\rm eff}$& ~log g~&~\vt~&~[Fe/H]~ & Rem \\
\hline
%                      T      g    vt    [Fe/H]     
  1& BS~16023-046 & 6364 &  4.50  & 1.3  & -2.97  & \\ 
  2& BS~16968-061 & 6035 &  3.75  & 1.5  & -3.05  & \\  
  3& BS~17570-063 & 6242 &  4.75  & 0.5  & -2.92  & \\  
  4& CS~22177-009 & 6257 &  4.50  & 1.2  & -3.10  & \\  
  5& CS~22888-031 & 6151 &  5.00  & 0.5  & -3.30  & \\  
  6& CS~22948-093 & 6356 &  4.25  & 1.2  & -3.30  & \\  
  7& CS~22953-037 & 6364 &  4.25  & 1.4  & -2.89  & \\  
  8& CS~22965-054 & 6089 &  3.75  & 1.4  & -3.04  & \\  
  9& CS~22966-011 & 6204 &  4.75  & 1.1  & -3.07  & \\  
 10& CS~29499-060 & 6318 &  4.00  & 1.5  & -2.70  & \\  
 11& CS~29506-007 & 6273 &  4.00  & 1.7  & -2.91  & \\  
 12& CS~29506-090 & 6303 &  4.25  & 1.4  & -2.83  & \\  
 13& CS~29518-020 & 6242 &  4.50  & 1.7  & -2.77  & \\ 
 14& CS~29518-043 & 6432 &  4.25  & 1.3  & -3.20  & \\  
 15& CS~29527-015 & 6242 &  4.00  & 1.6  & -3.55  & bin\\  
 16& CS~30301-024 & 6334 &  4.00  & 1.6  & -2.75  & \\  
 17& CS~30339-069 & 6242 &  4.00  & 1.3  & -3.08  & \\  
 18& CS~31061-032 & 6409 &  4.25  & 1.4  & -2.58  & \\  
\hline
 19& BS~16076-006 & 5199 &  3.00  & 1.4  & -3.81  & sg\\  
\hline 
\end {tabular}
\end {center}
\end {table}

\section{C, N, O abundances}
\subsection{Carbon}

The carbon abundance was determined by spectrum synthesis of the
$A^{2}\Delta - X^{2}\Pi$ band of CH (the G band).  Wavelengths of the
CH lines are from Luque and Crosley (\cite{LC99}); transition energies
are from the list of J{\o}rgensen et al.  (\cite{JLI96}); isotopic
shifts were computed using the best set of available molecular
constants.  The strongest lines of ${\rm ^{13}CH}$ at 423nm are
invisible in all of our stars, so the ${\rm ^{12}C/^{13}C}$ ratio
could not be measured.  In computing the total C abundance, we have
therefore assumed a solar ${\rm ^{12}C/^{13}C}$ ratio.

\begin {figure}
\begin {center}
\resizebox  {8.0cm}{4.5cm} 
{\includegraphics {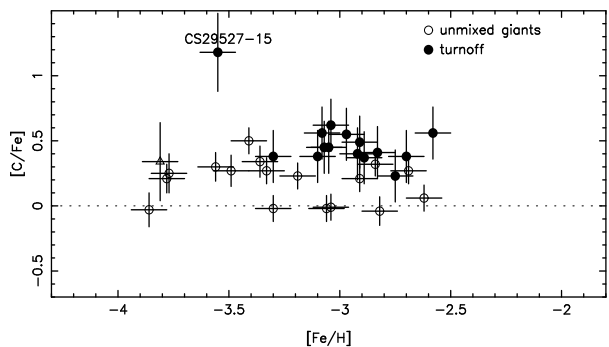} }
\caption{[C/Fe] ratios in our turnoff stars (black dots) and unmixed
giants (open circles).  The grey triangle shows the subgiant
BS~16076-006.}
\label {carbon}
\end {center}
\end {figure}

In Fig.  \ref{carbon} we present the measured [C/Fe] values in our
dwarf stars and compare them to values for our unmixed giants from
Paper V. In this figure we have omitted the mixed giants, located
above the bump, since we have shown (First Stars VI and IX) that the
abundances of C and N in the atmospheres of these stars are strongly
affected by mixing and thus are not good diagnostics of their initial
chemical compositions.

The mean [C/Fe] value for the turnoff stars is $\rm \overline{[C/Fe]}
= 0.45 \pm 0.10$ (s.d.), but $\rm \overline{[C/Fe]} = 0.19 \pm 0.16$ (s.d.)
for the giants.  Thus, we find a moderately significant difference between 
the C abundances in the giants and the turnoff stars (Fig.\ref{carbon}).  
We discuss the possible causes of this discrepancy in Sec.\ref{disc:3d}.
The mean [C/Fe] has been computed excluding
the binary turnoff star CS~29527-015, which  
appears to be quite carbon rich
(Fig.  \ref{carbon}).

%FIGURE 2

\subsection{Nitrogen}

Generally, the NH (and CN) bands are not visible in the spectra of EMP
turnoff stars (the stars are too hot), so N abundances can only be
measured in strongly N-enhanced stars (First Stars X).  The subgiant
BS~16076-006 exhibits a weak NH band, however, and we find [N/Fe]=
+0.29 for this star, taking into account the correction of $-0.4$ dex
derived in Paper VI.

\begin {figure}
\begin {center}
\resizebox  {8.0cm}{4.5cm} 
{\includegraphics {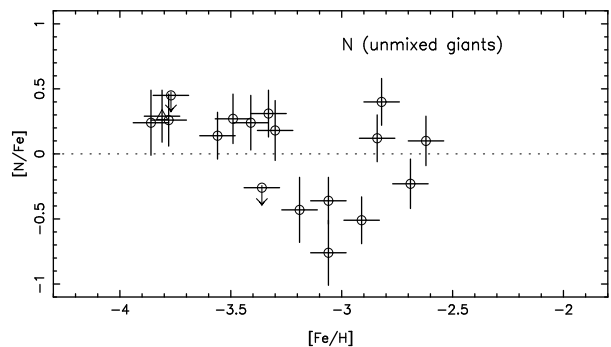} }
\caption{[N/Fe] in our sample of unmixed giants.  The 
triangle at [Fe/H]= --3.81 shows the subgiant BS~16076-006.}
\label {azote}
\end {center}
\end {figure}

Figure \ref{azote} shows the measured [N/Fe] ratios for our sample of
``unmixed'' giants (Paper VI).  BS~16076-006 agrees with (and thus
supports) the high [N/Fe] values found in the giants at the 
lowest metallicities.

%FIGURE 3

%TABLE 2
\begin {table*}[t]
\caption {Abundance ratios for C and the $\alpha$ elements (the subgiant 
BS16076-06 is shown separately).}
\label {tababund}
\begin {center}
\begin{tabular}{rll@{ ~}c@{ }c@{ } c@{ }c@{ }c@{ }c@{ }c@{ }c@{ }c@{ }c@{ }c@{ }c@{ }c@{ }c@{ }c@{ }c@{ }c@{ }c@{ }c@{ }c@{ }c@{ }c@{ }c }
\hline 
\hline
    &  Star       & [Fe/H]& [C/Fe]&$\sigma$&~& [Mg/Fe]&$\sigma$&N&~& [Si/Fe]&N&~& [Ca/Fe]&$\sigma$&N&~& [Sc/Fe]&N&~&     [Ti/Fe]&$\sigma$&N\\
\hline                                                                                                        
    1 &BS~16023-046& -2.97& 0.55& 0.15&~   &0.06& 0.06& 7&~ &    -0.07& 1&~ &    0.29& 0.09& 10& ~ &     0.10& 1&~ &      0.36& 0.06& 16   \\
    2 &BS~16968-061& -3.05& 0.45& 0.15&~   &0.29& 0.06& 7&~ &     0.31& 1&~ &    0.37& 0.10& 12& ~ &     0.44& 1&~ &      0.38& 0.05& 20   \\
    3 &BS~17570-063& -2.92& 0.40& 0.15&~   &0.08& 0.06& 7&~ &     0.04& 1&~ &    0.29& 0.10& 11& ~ &     0.29& 1&~ &      0.45& 0.06& 16   \\
    4 &CS~22177-009& -3.10& 0.38& 0.15&~   &0.22& 0.06& 7&~ &     0.15& 1&~ &    0.27& 0.08&  9& ~ &     0.21& 1&~ &      0.27& 0.05& 15   \\
    5 &CS~22888-031& -3.30& 0.38& 0.15&~   &0.23& 0.10& 7&~ &     0.31& 1&~ &    0.31& 0.16&  8& ~ &     0.28& 1&~ &      0.39& 0.04& 11   \\
    6 &CS~22948-093& -3.30& \nod&  -  &~   &0.05& 0.05& 6&~ &    -0.13& 1&~ &    0.30& 0.12&  4& ~ &     0.35& 1&~ &      0.49& 0.11& 15   \\
    7 &CS~22953-037& -2.89& 0.37& 0.15&~   &0.36& 0.08& 7&~ &    -0.01& 1&~ &    0.24& 0.10&  9& ~ &     0.35& 1&~ &      0.26& 0.06& 17   \\
    8 &CS~22965-054& -3.04& 0.62& 0.15&~   &0.25& 0.07& 7&~ &    -0.02& 1&~ &    0.47& 0.16& 13& ~ &     0.16& 1&~ &      0.44& 0.14& 25   \\
    9 &CS~22966-011& -3.07& 0.45& 0.15&~   &0.21& 0.08& 7&~ &     0.27& 1&~ &    0.32& 0.14& 10& ~ &     0.21& 1&~ &      0.38& 0.07& 16   \\
    10&CS~29499-060& -2.70& 0.38& 0.15&~   &0.19& 0.06& 7&~ &     0.00& 1&~ &    0.28& 0.06& 13& ~ &     0.10& 1&~ &      0.50& 0.07& 27   \\
    11&CS~29506-007& -2.91& 0.49& 0.15&~   &0.28& 0.05& 7&~ &     0.17& 1&~ &    0.49& 0.07& 13& ~ &     0.36& 1&~ &      0.52& 0.08& 23   \\
    12&CS~29506-090& -2.83& 0.41& 0.15&~   &0.27& 0.06& 7&~ &     0.17& 1&~ &    0.46& 0.10& 13& ~ &     0.27& 1&~ &      0.47& 0.07& 20   \\
    13&CS~29518-020& -2.77& \nod&  -  &~   &0.06& 0.03& 3&~ &     \nod& 1&~ &    0.40& 0.22&  7& ~ &     \nod& 1&~ &      \nod&  -  &  -   \\
    14&CS~29518-043& -3.20& \nod&  -  &~   &0.19& 0.09& 7&~ &     0.01& 1&~ &    0.40& 0.11&  9& ~ &     0.41& 1&~ &      0.49& 0.03& 15   \\
    15&CS~29527-015& -3.55& 1.18& 0.15&~   &0.43& 0.08& 7&~ &     0.15& 1&~ &    0.36& 0.23&  4& ~ &     0.26& 1&~ &      0.35& 0.12& 10   \\
    16&CS~30301-024& -2.75& 0.23& 0.15&~   &0.28& 0.07& 7&~ &     0.17& 1&~ &    0.45& 0.08& 14& ~ &     0.20& 1&~ &      0.45& 0.12& 25   \\
    17&CS~30339-069& -3.08& 0.56& 0.15&~   &0.18& 0.03& 7&~ &    -0.12& 1&~ &    0.43& 0.13& 10& ~ &     0.17& 1&~ &      0.38& 0.09& 20   \\
    18&CS~31061-032& -2.58& 0.56& 0.15&~   &0.22& 0.06& 7&~ &     0.14& 1&~ &    0.40& 0.14& 15& ~ &     0.31& 1&~ &      0.45& 0.11& 25   \\
\hline  
    - &BS~16076-006& -3.81& 0.34& 0.10&~   &0.58& 0.05& 7&~ &     0.31& 1&~ &    0.39& 0.14& 10& ~ &     0.42& 1&~ &      0.34& 0.07& 17   \\
\hline   
\end {tabular}  
\end {center}  
\end {table*}   
\

%TABLE 3  
\begin {table*}[t]  
\caption {Abundance ratios for the iron-peak and neutron-capture 
elements.} 
\label{tababund1}  
\begin {center}  
\begin{tabular}{rll@{ ~}c@{ }c@{ } c@{ }c@{ }c@{ }c@{ }c@{ }c@{ }c@{ }c@{ }c@{ }c@{ }c@{ }c@{ }c@{ }c@{ }c@{ }c@{ }r@{ }c@{ }c@{ }c@{ }c@{ }c@{ }c@{ }c@{ }c@{ }c@{ }c@{ }c@{ }c }
\hline 
\hline
    &  Star       & [Fe/H]& [Cr/Fe]&$\sigma$&N&&  [Mn/Fe]*&$\sigma$&N&& [Co/Fe]&$\sigma$&N && [Ni/Fe]&$\sigma$&N && [Zn/Fe]\hfill& N&& [Sr/Fe]& N&& [Ba/Fe]&N &~ \\
\hline
    1 &BS~16023-046& -2.97&  -0.12& 0.07& 5&~ & -0.55& 0.03& 3&~ &    0.28& 0.03& 2&~ &   -0.03& 0.15& 3&~ &   $<0.54$ & -&~ &   -0.18& 1&~ &    \nod& -&~  \\
    2 &BS~16968-061& -3.05&  -0.24& 0.06& 5&~ & -0.64& 0.00& 3&~ &    0.40& 0.04& 4&~ &    0.04& 0.07& 3&~ &   $<0.28$ & -&~ &   -0.57& 1&~ &    \nod& -&~  \\
    3 &BS~17570-063& -2.92&  -0.23& 0.12& 5&~ & -0.76& 0.01& 3&~ &    0.31& 0.08& 3&~ &   -0.07& 0.18& 3&~ &   $<0.41$ & -&~ &   -0.02& 1&~ &   -0.26& 1&~  \\
    4 &CS~22177-009& -3.10&  -0.22& 0.04& 5&~ & -0.57& 0.05& 3&~ &    0.37& 0.08& 3&~ &    0.02& 0.01& 2&~ &   $<0.37$ & -&~ &   -0.36& 1&~ &    \nod& -&~  \\
    5 &CS~22888-031& -3.30&  -0.28& 0.09& 4&~ & -0.74& 0.00& 2&~ &    0.57& 0.11& 3&~ &    0.08& 0.08& 2&~ &   \nod & -&~ &    0.18& 1&~ &    \nod& -&~  \\
    6 &CS~22948-093& -3.30&  -0.21& 0.08& 3&~ & -0.69& 0.00& 2&~ &    0.50&   - & 1&~ &   -0.01& 0.04& 2&~ &   $<0.82$ & -&~ &   -0.16& 1&~ &   -0.23& 1&~  \\
    7 &CS~22953-037& -2.89&  -0.32& 0.05& 5&~ & -0.78& 0.03& 3&~ &    0.39& 0.13& 3&~ &    0.04& 0.11& 3&~ &   $<0.39$ & -&~ &   -0.57& 1&~ &    \nod& -&~  \\
    8 &CS~22965-054& -3.04&  -0.16& 0.04& 5&~ & -0.51& 0.02& 3&~ &    0.44& 0.21& 4&~ &    0.03& 0.07& 3&~ &   0.67 & 1&~ &   +0.31& 1&~ &    \nod& -&~  \\
    9 &CS~22966-011& -3.07&  -0.23& 0.03& 5&~ & -0.70& 0.00& 3&~ &    0.48& 0.12& 4&~ &    0.08& 0.06& 2&~ &   $<0.50$ & -&~ &    0.03& 1&~ &   -0.05& 1&~  \\
    10&CS~29499-060& -2.70&   0.01& 0.04& 6&~ & -0.28& 0.02& 3&~ &    0.36& 0.09& 4&~ &    0.19& 0.09& 3&~ &   0.73 & 1&~ &   -0.60& 1&~ &    \nod& -&~  \\
    11&CS~29506-007& -2.91&  -0.12& 0.05& 5&~ & -0.59& 0.01& 3&~ &    0.39& 0.03& 3&~ &    0.04& 0.08& 3&~ &   0.71 & 1&~ &    0.16& 1&~ &    0.18& 1&~  \\
    12&CS~29506-090& -2.83&  -0.16& 0.06& 5&~ & -0.62& 0.02& 3&~ &    0.45& 0.11& 4&~ &    0.04& 0.12& 3&~ &   0.66 & 1&~ &    0.36& 1&~ &   -0.35& 1&~  \\
    13&CS~29518-020& -2.77&  -0.18& 0.05& 2&~ &  \nod& \nod&\nod&~ &  \nod&\nod & \nod&~ & 0.04&  \nod & 1&~ &   $<0.33$ & -&~ &    \nod& -&~ &    \nod& -&~  \\
    14&CS~29518-043& -3.20&  -0.20& 0.08& 4&~ & -0.64& 0.00& 2&~ &    0.57&   - & 1&~ &    0.07& 0.01& 2&~ &   $<0.68$ & -&~ &    0.08& 1&~ &    \nod& -&~  \\
    15&CS~29527-015& -3.55&  -0.21& 0.15& 4&~ & -0.66&   - & 1&~ &    0.70&   - & 1&~ &   -0.09& 0.05& 2&~ &   $<0.98$ & -&~ &    0.34& 1&~ &    \nod& -&~  \\
    16&CS~30301-024& -2.75&  -0.16& 0.06& 5&~ & -0.59& 0.01& 3&~ &    0.30& 0.11& 4&~ &    0.02& 0.04& 3&~ &   0.55 & 1&~ &   -0.32& 1&~ &   -0.28& 1&~  \\
    17&CS~30339-069& -3.08&  -0.24& 0.06& 5&~ & -0.71& 0.00& 3&~ &    0.33& 0.05& 2&~ &   -0.01& 0.17& 3&~ &   $<0.47$ & -&~ &   -0.10& 1&~ &    \nod& -&~  \\
    18&CS~31061-032& -2.58&  -0.10& 0.16& 6&~ & -0.51& 0.02& 3&~ &    0.38& 0.15& 4&~ &    0.03& 0.05& 3&~ &   0.40 & 1&~ &    0.21& 1&~ &   -0.40& 1&~  \\
\hline
     -&BS~16076-006& -3.81&  -0.41& 0.16& 6&~ & -0.93& 0.10& 3&~ &    0.39& 0.05& 4&~ &   -0.05& 0.04& 3&~ &   \nod & -&~ &$\le$-1.59& 1&~ &$\le$-1.0 & 1&~  \\
\hline 
\end {tabular}
\end {center}
* [Mn/Fe] has been determined only from the lines of the resonance triplet.
\end {table*}

\subsection{Oxygen}

We have not been able to measure O abundances for any of our dwarf
stars.  The [\ion{O}{I}] line at 630.03 nm, which we used for giants,
is too weak, as is the permitted \ion{O}{i} triplet at 770 nm, given
the S/N we achieve in this spectral region.  Only for the dwarf binary
system CS~22876-032 have we been able to measure O abundances, using
the OH lines in the UV (Gonz\`alez Hern\`andez et al.  \cite{jonay};
Paper XII), and these are compatible with the O abundances measured in
giants.

Our spectra of the dwarfs discussed in the present paper do not cover
the OH lines in the UV. The success in the case of CS~22876-032
suggests that these lines probably offer the only option for measuring
O abundances in EMP dwarfs.

\section{The $\alpha$ elements: Mg, Si, Ca, Ti}

Fig. \ref{MgSiCaTi} presents the observed [$\alpha$/Fe] ratios in our EMP 
dwarf and giant samples. 

A priori, we expect to find the same mean abundance for these elements
in dwarfs and in giants, and this is what we see for Ca.  
However, the mean [Mg/Fe] and [Si/Fe] ratios are $\sim$~0.2 and 
0.3 dex lower in the EMP
dwarfs than in the giants, while the mean abundance of [Ti/Fe] ratio
is {\it higher} in the dwarfs by about 0.2 dex.  What are the possible causes
of these differences?

\subsection{ Magnesium} \label{mag}

In Fig. \ref{MgSiCaTi}, the Mg abundance for the giant stars has 
been derived from a full fit to the profiles of the Mg lines, in contrast 
to the results given by Cayrel et al.  (\cite{CDS04}, Paper V).  The
equivalent widths of the Mg lines are often quite large ($\rm
EW > 120mA$), and in Paper V  we underestimated the equivalent widths 
of these lines by neglecting the wings.  For the most Mg-poor stars 
in our sample the lines are weak and the difference negligible, but 
it is quite significant in most of our stars, with a mean systematic 
difference of about 0.15 dex.  
In the dwarfs, the abundance has been derived from profile fits to the
strongest lines (the lines at $\sim$383 nm, which are also located in the 
wings of a Balmer line); and from equivalent widths for the weak lines.  

\subsection{ Silicon}

In the cool giants the Si abundance is derived from a line at 410.3
nm.  This line (multiplet 2) is located in the wing of H$\delta$, and
the hydrogen line has been included in the computations.  There is
another line at 390.6 nm (multiplet 3), but in giants this line is
severely blended by CH lines.  In turnoff stars the line at 410.3 nm
is invisible, but the CH lines are weak enough that the line at 390.6
nm can be used.  Thus, in the end, only a single Si line (but not the
same one) could be used in both dwarfs and giants; a systematic error
in the log {\sl gf} of these lines could explain the observed
difference.  
Both lines are in fact measured in the subgiant star BS\,16076-006 and 
yield consistent Si abundances, but, given the uncertain atmospheric 
parameters of this star (see \ref{analysis}), a systematic error in 
log {\sl gf} cannot be ruled out. 
Our new [Si/Fe] ratios are in good agreement with the
value found from the same Si line by Cohen (\cite{CCM04}), also
for EMP turnoff stars.

\subsection{ Titanium}

The Ti~I lines are very weak in turnoff stars, so the Ti 
abundance can only be determined from the Ti~II lines.  About 15 Ti~II
lines could be used, and the internal error of the mean is very small
(less than 0.1 dex).  Fig.  \ref{MgSiCaTi} clearly shows higher
[Ti/Fe] ratios in the dwarfs than in the giants ($\rm \Delta [Ti/Fe] =
0.2$ dex).  Even if we use exactly the same lines in the giants as in
the dwarfs, we observe the same effect; thus, an error in log {\sl gf}
values cannot explain the difference.  On the other hand, to reduce
the derived [Ti/Fe] by 0.2 dex (keeping the same temperature) would
require changing log $g$ in the turnoff stars by about 1 dex, which is
quite incompatible with the ionisation equilibrium of the iron lines.

\section{The light odd-Z metals: Na, Al, K, and Sc}

% TABLE 4
\begin {table}[t]
\caption {Abundance uncertainties linked to stellar parameters.}
\label {errors}
\begin {center}
\begin {tabular}{lrrrr}
\hline
\multicolumn {2}{l}{CS~22177-009}\\
\multicolumn {5}{c}{A: T$_{\rm eff}$=6260~K, log g = 4.5, vt=1.3 
km s$^{-1}$}\\
\multicolumn {5}{c}{B: T$_{\rm eff}$=6260~K, log g = 4.4, vt=1.3 
km s$^{-1}$}\\
\multicolumn {5}{c}{C: T$_{\rm eff}$=6260~K, log g = 4.5, vt=1.1 
km s$^{-1}$}\\
\multicolumn {5}{c}{D: T$_{\rm eff}$=6160~K, log g = 4.5, vt=1.3 
km s$^{-1}$}\\
\multicolumn {5}{c}{E: T$_{\rm eff}$=6160~K, log g = 4.3, vt=1.2 
km s$^{-1}$}\\
\hline
El.   & $\Delta_{B-A} $ & $\Delta_{C-A} $& $\Delta_{D-A} $& 
$\Delta_{E-A} $\\
\hline
[Fe/H]      &-0.01 & 0.03 &-0.05 &-0.06\\
$[$Na I/Fe] & 0.02 &-0.02 &-0.01 & 0.01\\
$[$Mg I/Fe] & 0.03 &-0.01 &-0.02 & 0.00\\
$[$Al I/Fe] & 0.01 &-0.03 &-0.03 &-0.01\\
$[$Si I/Fe] & 0.03 & 0.01 &-0.03 & 0.02\\
$[$Ca I/Fe] & 0.01 &-0.02 & 0.00 & 0.01\\
$[$Sc II/Fe]&-0.02 &-0.02 & 0.00 &-0.05\\
$[$Ti I/Fe] & 0.01 &-0.03 &-0.03 &-0.03\\
$[$Ti II/Fe]&-0.02 &-0.01 & 0.01 &-0.03\\
$[$Cr I/Fe] & 0.01 &-0.02 &-0.03 &-0.02\\
$[$Mn I/Fe] & 0.01 &-0.02 &-0.04 &-0.03\\
$[$Fe I/Fe] & 0.02 & 0.01 &-0.03 & 0.01\\
$[$Fe II/Fe]&-0.03 &-0.02 & 0.03 &-0.02\\
$[$Co I/Fe] & 0.01 &-0.03 &-0.04 &-0.03\\
$[$Ni I/Fe] & 0.01 &-0.01 &-0.04 &-0.03\\
$[$Sr II/Fe]&-0.02 & 0.01 &-0.01 &-0.04\\
$[$Ba II/Fe]&-0.02 & 0.00 &-0.01 &-0.04\\
\hline
\end {tabular}
\end {center}
\end {table}

\subsection{ Sodium and Aluminium}

In both dwarf and giant EMP stars, Na and Al abundances can only be
derived from the resonance lines, which are very sensitive to NLTE
effects (Cayrel et al.  \cite{CDS04}).  The Na and Al abundances in
our two stellar samples have been derived using the NLTE line
formation theory by Andrievsky et al.  (\cite{ASK07}) and Andrievsky
et al.  (\cite{ASK08}) for Na and Al, respectively.  When NLTE effects
are taken into account, the [Na/Fe] and [Al/Fe] abundance ratios are
found to be constant and equal in the dwarfs and giants in the
interval $\rm -3.7<[Fe/H]<-2.5$ ([Na/Fe] = $-0.2$ and [Al/Fe] =
$-0.1$). This can be appreciated visually by looking at 
figure  7 of Andrievski et al. (\cite{ASK07})
and figure 3 of Andrievski et al. (\cite{ASK08}). 

%\afterpage{\clearpage%FIGURE 4
\begin {figure}
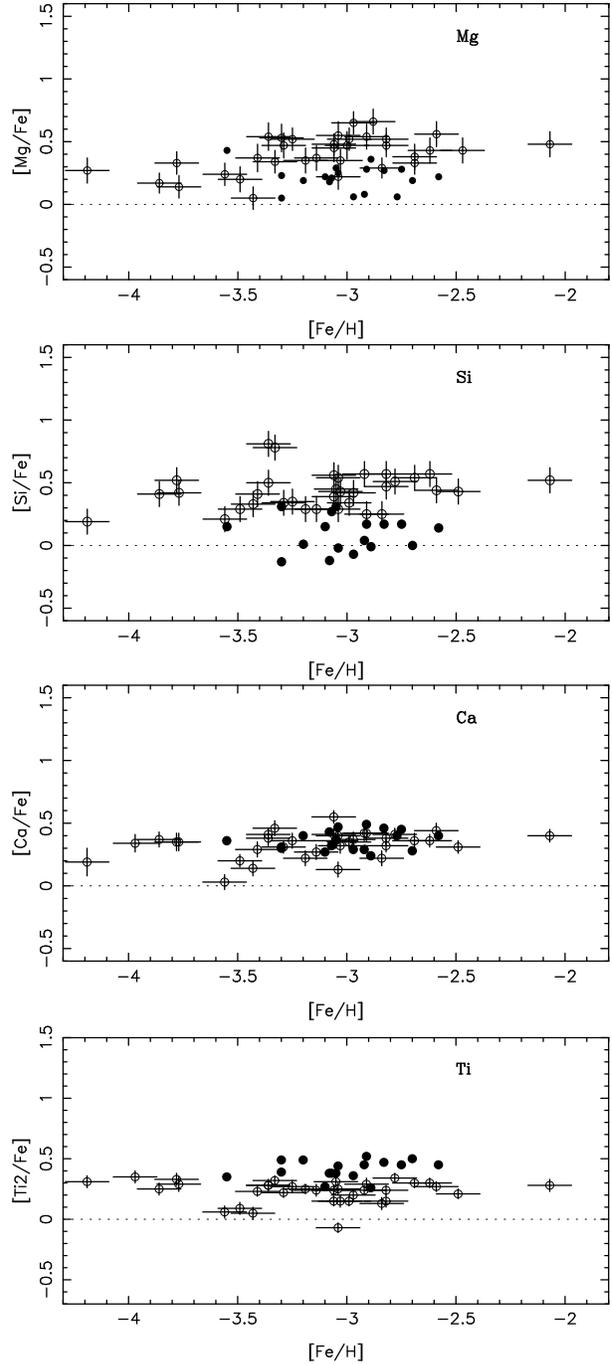

\begin {center}
\resizebox  {8.0cm}{4.5cm} 
{\includegraphics {10610f4a.ps} }
\resizebox  {8.0cm}{4.5cm} 
{\includegraphics {10610f4b.ps} }
\resizebox  {8.0cm}{4.5cm} 
{\includegraphics {10610f4c.ps} }
\resizebox  {8.0cm}{4.5cm} 
{\includegraphics {10610f4d.ps} }
\caption{[Mg/Fe], [Si/Fe], [Ca/Fe], and [Ti/Fe] in our program stars. 
Symbols as in Fig. \ref{carbon}.}
\label {MgSiCaTi}
\end {center}
\end {figure}
%}

\subsection{ Potassium and Scandium}

The K lines are very weak in our EMP turnoff stars, and [K/Fe] could not 
be determined.

\begin {figure}
\begin {center}
\resizebox  {8.0cm}{4.5cm} 
{\includegraphics {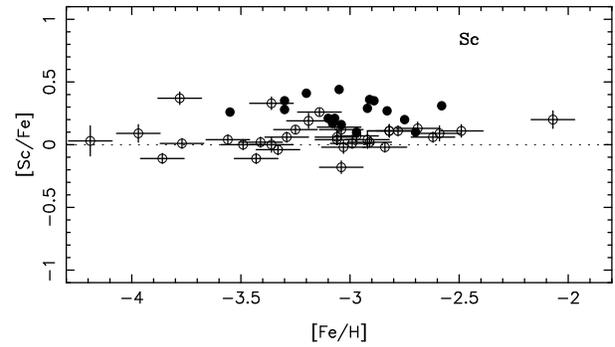} }
\caption{[Sc/Fe] in the programme stars. Symbols as in Fig. \ref{carbon}.}
\label {Sc}
\end {center}
\end {figure}

The Sc abundance in the dwarf stars has been measured from the Sc~II line 
at 424.6 nm. In giants, 7 Sc lines could be used, and the 
scatter in the abundances from individual lines is very small (below 
0.1 dex). There is a systematic difference of about 0.2 dex between the 
Sc abundances in the giants and the dwarfs (Fig. \ref{Sc}).

\section{Iron-peak elements}

\subsection{Chromium, Cobalt, and Nickel\label{CrCoNi}}

Fig.  \ref{ironpeak} shows the [Cr/Fe], [Co/Fe], and [Ni/Fe] ratios
for our dwarf and giant samples.  There is rather good agreement for
Ni, but [Cr/Fe] and [Co/Fe] are about 0.2 dex higher in the dwarfs
than in the giants.  Recently Lai et al.  (2008) have measured the
chromium abundance in a sample of giants and turnoff stars in the same
range of metallicity.  The same shift appears between their giants and
turnoff stars (Fig.  \ref{CrLai}).

Lai et al.  also found an offset between the abundances derived from
Cr~I and Cr~II. Cr~II can only be measured in giants, and only a
single Cr~II line ($\rm \lambda=455.865nm$) appears at the edge of our
blue spectra, but the same offset as observed by Lai et al.  is
clearly visible in our data (Figure \ref{Cr1-2}).

The discrepancy between Cr~I and Cr~II, and between giants and turnoff
stars, may point to non-LTE effects.  The main Cr~I lines are
resonance lines.  Unfortunately no precise structure model for the Cr
atom exists, so it is not possible to explore this hypothesis at
present.  If significant NLTE effects were confirmed, the most
reliable abundances should be those from the Cr~II line, suggesting that
$\rm [Cr/Fe]\approx +0.1$ at low metallicity.

Nissen \& Schuster (\cite {NS97}) found a close correlation between
the abundances of Na and Ni in the interval $\rm -0.7<[Fe/H]<-1.3$.
To explain this correlation, it has been suggested that the production
of $\rm ^{58}Ni$ during an SN~II event depends on the neutron excess,
which itself depends mainly on the amount of $\rm ^{23}Na$ produced
during hydrostatic carbon burning.  However, this correlation is not
observed in our sample (Fig.  \ref{nani}).  In fact, [Ni/Fe] and
[Na/Fe] have the same mean value in turnoff stars as in unmixed giants
([Ni/Fe] = 0.0 and [Na/Fe] = $-0.2$).  [Na/Fe] is larger in several of
the mixed giants, but this is due to mixing with the H-burning shell
in layers that are sufficiently deep to bring products of the Ne-Na
cycle to the surface (see Andrievsky et al.  \cite{ASK07}).

\subsection{Manganese}

The Mn abundances have been derived by fitting synthetic spectra to the 
observations, taking into account the hyperfine structure of the lines.
We noted in Paper V that, in the giant stars, Mn abundances determined 
from the resonance lines were lower than those from the lines of higher 
excitation potential by about 0.4 dex. At this stage, we prefer the 
abundances from the high-excitation lines, because the resonance lines 
are more susceptible to non-LTE effects. However, in the five most 
metal-poor giants only the resonance triplet is detected, so for these 
stars the Mn abundance was determined from the triplet and corrected by 
the adopted 0.4\,dex offset.

For most of the turnoff stars analysed here, again only the resonance 
triplet can be detected. In Fig. \ref{manga}a the Mn abundances from these 
lines have been systematically increased by 0.4 dex, while in Fig. 
\ref{manga}b [Mn/Fe] is derived from the resonance triplet profiles in 
all the stars and plotted without any correction.

In both cases we find a systematic abundance difference of about 0.2 dex 
between the giants and the dwarfs.

\subsection{Zinc}

Zinc cannot be unambiguously assigned to the iron-peak category, since
it may be formed by $\alpha$-rich freeze-out and neutron capture as
well as by burning in nuclear statistical equilibrium.  In our sample,
the only usable line is the strongest \ion{Zn}{i} line of Mult.  2, at
481\, nm.  The line is very weak in all our stars, and we only
consider it reliably detected and provide a measurement when the
equivalent width is larger than 0.35 pm.  Thus, Table \ref{tababund1}
gives only six measurements and eleven upper limits; for two stars, the
spectrum was affected by a defect, and it is not even possible to
provide an upper limit.

%FIGURE 5

Figure \ref{zinc} shows [Zn/Fe] versus [Fe/H]; upper limits are shown
as downward arrows and the giant stars from Cayrel et al.
(\cite{CDS04}) as open circles.  The upper limits are consistent with
the trend defined by the giant stars, but the six actual measurements
appear to define a similar trend, shifted upwards by about 0.4 dex.
This could be another example of the dwarf/giant discrepancy found for
some other elements.

  Since the majority of our [Zn/Fe] data are upper limits,
 we use survival statistics to analyse them.  The giant stars with
 [Fe/H]$\ge -3.0$ show a constant level of [Zn/Fe]$=+0.199\pm0.080$.
 We selected the dwarf stars in the same metallicity range and used
 {\tt asurv Rev 1.2}
 \footnote{http://astrostatistics.psu.edu/statcodes/asurv} (Lavalley
 et al.  \cite{baas}) to compute the Kaplan-Meier statistics, as
 described in Feigelson \& Nelson (\cite{FN}).  The mean is $+0.491\pm
 0.055$; since the lowest point is an upper limit, it has been changed
 to a detection to compute the Kaplan-Meier statistics, which implies
 that this mean value is biased.  The comparison of the two mean
 values, for giants and dwarfs suggests that they are only marginally
 consistent:  The 75th percentile of the [Zn/Fe] values for dwarfs 
(+0.223) corresponds to 
 the mean value for giants.  Changing the upper limits to $2
 \sigma$ or $3 \sigma$ would push the mean value for dwarfs even
 higher, thus making the values of dwarfs and giants even more
 inconsistent.

We used the generalized version of Kendall's $\tau$ (Brown et al.
\cite{brown}), as described in Isobe et al.  (\cite{isobe}), to check if
there is support for a correlation between [Fe/H] and [Zn/Fe] for the
dwarf stars.  The sample is composed of 6 detections and 11 upper
limits.  The probability of correlation is 91.3\%, so there is a
hint of a correlation, but no conclusive evidence.

For Zn it appears unlikely that  the giant/dwarf discrepancy is
due to NLTE effects: Takeda et al.  (\cite{takeda}) computed NLTE
corrections for the \ion{Zn}{i} line at 481\, nm; the corrections are
small and {\em negative} for metal-poor giants, {\em positive} for
metal-poor TO stars.  Thus, if we applied these corrections to our
sample, the discrepancy would increase from 0.2 to $\sim 0.4$ dex.

It is surprising that several stars display upper limits which are
{\em lower} than the [Zn/Fe] ratios found in other stars of similar
metallicity, suggesting a real cosmic scatter in the Zn abundance.  It
is interesting to note that while for giant stars Lai et al.
(\cite{LBJ08}) are in good agreement with our determinations, the two
dwarf stars for which they have Zn measurements appear to be in line
with the measurements of giants.  This may give further support to the
idea of a cosmic scatter of Zn abundances, or to the existence of a
Zn-rich population.

However, it should be kept in mind that the available Zn lines are all
very weak (detections are about 0.4 pm, upper limits 0.1--0.2 pm), and
the data should not be overinterpreted.  We have, perhaps somewhat
na\"ively, placed the upper limit at the measured value for all stars
below our chosen threshold.  Had we decided to put the upper limit at
$3\sigma$ above the measured EW, all the upper limits would move up
among the measurements or beyond, and there would be no hint of any
scatter in the Zn abundance.   From the point of view of survival
statistics, the fact that the standard deviation from the mean is
small, compared to observational errors, does not support the
presence of a real dispersion. The question of a scatter in Zn abundance
in EMP dwarf stars clearly needs further study, if possible based on
different lines.

%FIGURE 6
\begin {figure}
\begin {center}
\resizebox  {8.0cm}{4.5cm} 
{\includegraphics {10610f6a.ps} }
\resizebox  {8.0cm}{4.5cm} 
{\includegraphics {10610f6b.ps} }
\resizebox  {8.0cm}{4.5cm} 
{\includegraphics {10610f6c.ps} }
\caption{[Cr/Fe], [Mn/Fe], [Co/Fe], and [Ni/Fe] in the program stars.  
Symbols as in Fig. \ref{carbon}.}
\label {ironpeak}
\end {center}
\end {figure}

%FIGURE 7
\begin {figure}
\begin {center}
\resizebox  {8.0cm}{4.5cm} 
{\includegraphics {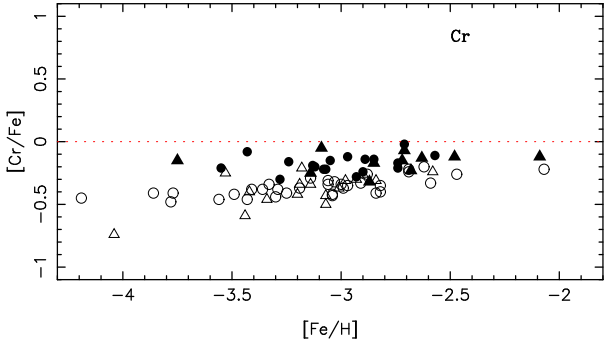} }
\caption{[Cr/Fe] in our stars (circles) compared to those of Lai et al. 
(2008, triangles). Filled symbols: turnoff stars; open symbols: giants.}
\label {CrLai}
\end {center}
\end {figure}

\section{Neutron-capture elements}

Very few neutron-capture elements can be measured in turnoff stars,
because their lines are generally very weak.  We could, however,
measure Sr abundances from the blue resonance line of Sr~II, and
sometimes also Ba abundances from the Ba II line at 455.4 nm.  The Ba
line is generally weak (about 0.5 pm) and located at the very end of
the blue spectrum, where the noise is higher.  As Fig.  \ref{heavy}
shows, we find good agreement between dwarfs and giants, although the
star-to-star scatter is very large, as has already been observed for
the giant stars.  In Fig.  \ref{srba} we show the [Sr/Ba] ratio as a
function of [Ba/H].  As already noticed in Paper VIII (Fig.  15), the
scatter in this plane is greatly reduced.  The dwarf stars appear to
behave exactly in the same way as giants.

We have recently studied the Ba abundance in dwarfs and giants taking 
into account non-LTE effects (Andrievsky et al.  \cite{ASK09}), but 
after correcting for NLTE the general behaviour of this element remains 
the same.

\section{Comparison with other investigations.}

Several other groups have now published detailed analyses of EMP stars
similar to our own, and it is interesting to compare their results to
ours.  We focus on the results of the 0Z project (Cohen et al.
\cite{CCM04,CCM08}) and Lai et al.  (\cite{LBJ08}).  The details of
the comparison are provided in appendices \ref{0Zcomp}, \ref{laicomp}
and \ref{BScomp}.  The final conclusion of this comparison is that there is
excellent agreement between the three groups, and the small differences
can be understood in terms of differences in the adopted atmospheric
parameters, model atmospheres or line selection. The MARCS model atmospheres 
used by us agree with the ATLAS
non-overshooting models adopted by Lai et al.  (\cite{LBJ08}), and both
yield abundances which are about 0.1\, dex lower than the ATLAS
overshooting models adopted by the 0Z project.

%FIGURE 8
\begin {figure}[t!]
\begin {center}
\resizebox  {8.0cm}{4.5cm} 
{\includegraphics {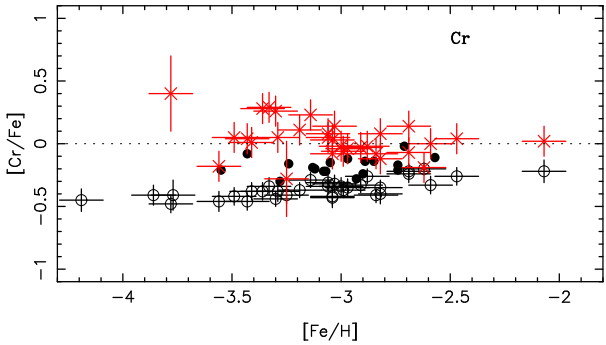} }
\caption{[Cr/Fe] in our stars from Cr~I lines 
(open circles: giants, filled circles: turnoff stars), and 
from the single Cr~II line (crosses; giants only). The large offset cannot 
be explained by measurement errors.
}
\label {Cr1-2}
\end {center}
\end {figure}

\section{Discussion\label{disc:3d}}

For most elements, the overall abundance trends defined by dwarfs and
giants show good agreement. For example, [Ti/Fe] is constant at low 
metallicity in both giants and dwarfs, and [Mn/Fe] decreases with 
metallicity in both giants and dwarfs.

However, some elements show systematic shifts in [X/Fe] between turnoff 
stars and giants of the same metallicity.  Generally,
$\rm [X/Fe]_{dwarfs} - [X/Fe]_{giants} \approx +0.2$ dex, except for Mg and 
Si, which show a negative shift. Also, [Cr/Fe] appears to be flat in the 
dwarfs, but displays a significant slope for the giants. 
It is difficult to explain these shifts
by systematic errors in the models (error in temperature or in
gravity) because the effects on the abundance of all the elements are
very similar (see Table \ref {errors}), so the ratios [X/Fe] are
little affected.

These differences 
are rather puzzling because, except for C, N, and possibly Na, the
chemical composition of the giant stars should be unaltered since the
star formed, so one would expect that the abundances in giants should
match those in dwarfs at any given metallicity.  The discrepancy we
find is most likely due to shortcomings in our analysis, but we do not
know whether we should trust the derived results for giants or dwarfs
(or perhaps neither!).

In the following we discuss the two main simplifications of standard 
model atmospheres, the neglect of effects of granulation (``3D effects'' 
for short) and deviations from local thermodynamic equilibrium (NLTE), 
as possible causes of the observed discrepancies.

\subsection{Granulation (3D) effects}

It is well known that hydrodynamical simulations (``3D models'')
predict much cooler temperatures in the outer layers of metal-poor 
stars than 1D models (Asplund et al. \cite{asp99}, 
Collet et al. \cite{collet}, Caffau \& Ludwig \cite{CL07},
Gonz\'alez Hern\'andez et al. \cite{jonay}, Paper XI). The effect 
is more pronounced for dwarfs than for giants. The species most 
affected by this difference are clearly those which predominantly 
reside in such cool layers, most notably the diatomic molecules such 
as CH and NH. Since one of the most striking differences between 
dwarfs and giants is in fact the C abundance, which we derive from 
CH lines, we decided to investigate the effects of granulation in
more detail.

To accomplish this, we used the two \cobold\ (Freytag et al.
\cite{freytag02}, Wedemeyer et al.  \cite{wedemeyer04}) 3D models
described in Paper XI (\Teff/log g/[Fe/H]: 6550/4.50/--3.0 and
5920/4.50/--3.0).  Unfortunately, we do not yet have any fully relaxed
models for giant stars, so we decided to use a representative snapshot
of a 3D simulation of a giant close to relaxation (\Teff/log g/[Fe/H]:
4880/2.00/--3.0).
 
%FIGURE 9
\begin {figure}[b!]
\begin {center}
\resizebox  {7.0cm}{7.0cm} 
{\includegraphics {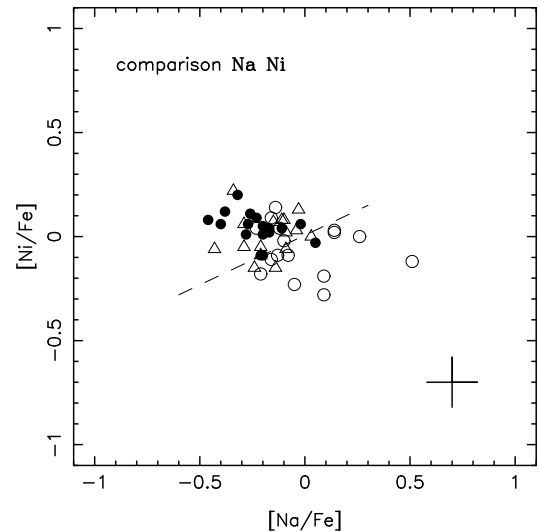} }
\caption{[Ni/Fe] vs. [Na/Fe] in dwarfs and giants; symbols as in Fig. 
\ref{carbon}.  The dashed line shows the correlation found by Nissen \& 
Schuster (\cite{NS97}) for $\rm -0.7<[Fe/H]<-1.3$ and corresponds to
the expected production ratios of Na and Ni in Type II supernovae. We
do not observe this correlation in our sample.  The few large [Na/Fe] 
values ($\rm [Na/Fe]>+0.1$) refer to some of the more extreme ``mixed''
giants discussed in Paper IX.
}
\label {nani}
\end {center}
\end {figure}

Table \ref{3dcor} lists the mean 3D corrections as defined by Caffau
\& Ludwig (\cite{CL07}) for the three models described above.  The
sense of the correction is always 3D-1D. Approximating the 3D
correction for the G-band as the average for just 4 lines is
admittedly somewhat crude, but should provide a reliable
order-of-magnitude estimate for the effect.
  
%FIGURE 10
\afterpage{\clearpage
\begin {figure}[b!]
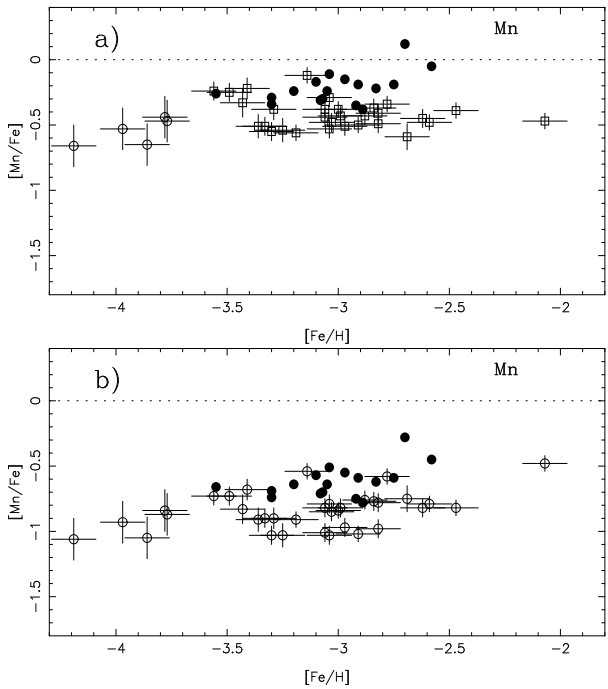
 \begin {center} \resizebox {8.0cm}{4.5cm}
{\includegraphics {10610f10a.ps} } \resizebox {8.0cm}{4.5cm}
{\includegraphics {10610f10b.ps} } \caption{[Mn/Fe] ratios for
our dwarf and giant stars.  In panel a), [Mn/Fe] in most of the giant
stars is determined only from lines with excitation potential $>2$eV
(open squares).  In the turnoff stars (black dots) and the five most
metal-poor giants (open circles), only the resonance triplet of Mn is
usable; [Mn/Fe] is then derived from these lines, corrected by +0.4
dex (see text).  In panel b), [Mn/Fe] is determined from the resonance
lines for all stars (giants and dwarfs), without any correction.  In
both cases, the mean [Mn/Fe] ratio is offset by $\sim$0.2 dex between
the giant and dwarf stars.  } \label {manga} \end {center} 
\end {figure}
}

For the C abundance, the effect is quite prominent for dwarfs.  The
magnitude of the correction is such that, if applied, the discrepancy
in [C/Fe] between dwarfs and giants would be somewhat reduced (from
0.27 dex to 0.13 dex), but with the opposite sign, the dwarfs now
showing a slightly lower C abundance.  Given the crudeness of our 3D
computations, we cannot claim with certainty that 3D effects will
explain the discrepancy.  To the extent that our order-of-magnitude
estimates are reliable, it is possible that more accurate computations
with a larger set of parameters, encompassing the full range of our
dwarf and giant stars, would yield [C/Fe]$\sim 0.2$ for both dwarfs
and giants.

For the giant model, our computed correction for C is a factor of two
smaller than the results of Collet et al.  (\cite{collet}).  Also, for
Fe, our corrections are considerably smaller than found by Collet et
al., especially for the resonance line.

The issue clearly requires further investigation, which we will
undertake when we have several fully relaxed 3D models of giants.  A
detailed discussion is therefore premature.  At present it is unclear
whether the different results we find are due to some fundamental
difference between the 3D codes: \cobold ~(Freytag et al.
\cite{freytag02}, Wedemeyer et al.  \cite{wedemeyer04}) in our case,
and the Stein \& Nordlund (\cite{SN98}) code in the case of Collet et
al.  (\cite{collet}), or simply to the choice of a MARCS model as the
1D reference by Collet et al.  (\cite{collet}).

We performed spectrum synthesis computations, using 
\linfor\footnote{http://www.aip.de/~mst/Linfor3D/linfor\_3D\_manual.pdf}
to estimate 3D corrections for a few selected lines, listed in 
Table \ref{tab3dlin}. This is not meant to substitute for a full 3D 
investigation of the sample, but should provide an indication
whether the differences found between giants and dwarfs might vanish 
if suitable 3D models were used.

For Si, Co, and Zn the predicted corrections are the same for giants 
and dwarfs, so the discrepancy for these elements should not be due 
to granulation effects. For Sc and Ti, however, the differences go in 
the direction of increasing the discrepancy between dwarfs and giants.

For Mn and Cr, the difference in correction between dwarfs and giants
is such as to exactly cancel the discrepancies.  We caution, however,
that the corrections listed in Table \ref{3dcor} are the average of
those for the resonance and high excitation lines.  The difference in
correction between the two lines is smaller for the giant model (0.1
dex for Cr, 0.3 dex for Mn) than for the dwarf models (0.4--0.5 dex
for Cr, 0.7 dex for Mn).  This difference is still somewhat
problematic, however, in the sense that while the 1D analysis achieved
a good excitation equilibrium for Cr, an analysis based on the 3D
atmospheres does not.  This suggests that the temperature scale
appropriate for 3D models may in fact be different from those adopted
in this paper and by Cayrel et al.  (\cite{CDS04}).  As mentioned
above, a 1D analysis implies a Mn abundance about 0.4 dex {\em lower}
for the resonance lines than for the high-excitation lines, and the 3D
corrections for Mn {\em increase} this difference, up to 1.1 dex.

\begin{table}
\centering
\caption{\label{tab3dlin}Lines used to test the granulation effects.}
\begin{tabular}{rcc}
\hline
\hline
Species & $\lambda$  & $\chi$ \\
        &  nm        & eV   \\
\hline
CH           & 430.0317 & 0.00 \\
CH           & 430.0587 & 0.36 \\
CH           & 430.1072 & 1.44 \\
CH           & 430.1135 & 0.31 \\
\ion{Si}{i}  & 390.5523 & 1.91 \\ 
\ion{Si}{i}  & 410.2916 & 1.91 \\ 
\ion{Sc}{ii} & 424.6822 & 0.31 \\
\ion{Ti}{ii} & 376.1323 & 0.57 \\
\ion{Ti}{ii} & 391.3468 & 1.12 \\ 
\ion{Cr}{i}  & 425.4332 & 0.00  \\
\ion{Cr}{i}  & 520.8419 & 0.94 \\  
\ion{Mn}{i}  & 403.0753 & 0.00 \\
\ion{Mn}{i}  & 404.1355 & 2.11 \\ 
\ion{Fe}{i}  & 382.4444 & 0.00 \\
\ion{Fe}{i}  & 400.5242 & 1.56 \\
\ion{Fe}{i}  & 418.7795 & 2.42 \\
\ion{Fe}{i}  & 422.7427 & 3.33 \\
\ion{Co}{i}  & 384.5461 & 0.92 \\ 
\ion{Zn}{i}  & 481.0528 & 4.08 \\
\hline
\end{tabular}
\end{table}

\begin{table}
\caption{Mean 3D corrections for selected elements.\label{3dcor}}
\centering
\begin{tabular}{rccc}
\hline\hline
        & \multispan{3} model \\
        &  4880/2.00/--3.0 & 5920/4.50/--3.0 & 6550/4.50/--3.0 \\
\hline
 \relax [C/H]   & $ -0.1 $         &$ -0.5 $         &$ -0.6$\\
 \relax [Si/H]  & $ -0.1 $         &$ -0.1 $         &$ -0.2$\\
 \relax [Sc/H]  & $ -0.2 $         &$ -0.1 $         &$ -0.1$\\
 \relax [Ti/H]  & $ -0.1 $         &$  0.0 $         &$  0.0$\\
 \relax [Cr/H]  & $ -0.3 $         &$ -0.6 $         &$ -0.5$\\
 \relax [Mn/H]  & $ -0.3 $         &$ -0.5 $         &$ -0.5$\\   
 \relax [Fe/H]  & $ -0.2 $         &$ -0.2 $         &$ -0.3$\\
 \relax [Co/H]  & $ -0.3 $         &$ -0.3 $         &$ -0.4$\\
 \relax [Zn/H]  & $ +0.1 $         &$ +0.1 $         &$ +0.1$\\ 
\hline
 \relax [C/Fe]  & $ +0.1 $         &$-0.3  $         &$ -0.3$\\
 \relax [Si/Fe] & $ +0.1 $         &$+0.1  $         &$ +0.1$\\
 \relax [Sc/Fe] & $  0.0 $         &$+0.1  $         &$ +0.2$\\
 \relax [Ti/Fe] & $ +0.1 $         &$+0.2  $         &$ +0.3$\\
 \relax [Cr/Fe] & $ -0.1 $         &$-0.4  $         &$ -0.2$\\
 \relax [Mn/Fe] & $ -0.1 $         &$-0.3  $         &$ -0.2$\\
 \relax [Co/Fe] & $ -0.1 $         &$-0.1  $         &$ -0.1$\\ 	
 \relax [Zn/Fe] & $ +0.3 $         &$+0.3  $         &$ +0.4$\\ 	
\hline

\end{tabular}
\end{table}

It is unlikely that the use of 3D models will bring the abundances in
giants and dwarfs into agreement for all elements, although it may be
possible for a few (most likely C, Cr, and Mn).  However, a full
re-analysis based on 3D models, including a redetermination of the
atmospheric parameters, is needed before reaching a firm conclusion on
this point.  For the time being, since the predicted 3D corrections
are always smaller for our giant model than for the dwarf models, we
consider the 1D abundances for giants to be more reliable than for the
dwarfs.

\subsection{Deviations from local thermodynamic equilibrium.}

The analysis in this paper and in Cayrel et al.  (\cite{CDS04}) is
based on the assumption of local thermodynamic equilibrium (LTE), both
in the computation of the model atmospheres and in the line transfer
computations.  For Na and Al, results based on NLTE line transfer
computations have been presented in Andrievsky et al.  (\cite{ASK07,
ASK08}).  For both elements, the LTE computations implied a
discrepancy between dwarfs and giants, while the NLTE computations
provided consistent abundances between the two sets of stars.  In the
case of Na, the NLTE corrections are not very different for dwarfs or
giant models for lines of a given equivalent width, but the correction
depends strongly on the equivalent width.  The giant stars, which are
cooler, have larger equivalent widths and larger NLTE corrections.  In
this case the LTE abundances of dwarfs are to be considered more
reliable than those of giants.

The result cannot be generalized, however, so detailed NLTE
computations should be carried out for all the elements for which we
find a discrepancy between dwarfs and giants.  Also, from the point of
view of departures from NLTE, one cannot {\em a priori} assume that
the departures are larger for the stronger lines (i.e. for giants),
although this is often the case.
 
Accordingly, except for the two elements Na and Al for which we
already have NLTE computations, we cannot at present say whether
accounting for NLTE effects could remove the discrepancy between
dwarfs and giants.  Computations of the NLTE abundance of Mg
are under way.

\subsection{Could the dwarf/giant discrepancy be real ?}

For C, the difference in [C/Fe] between dwarfs and giants might
represent the effect of the first dredge-up, which could be
responsible for a decrease of the C abundance due to a first mixing
with the H-burning layer, where C is transformed into N. For the other
elements we see no possible nucleosynthetic origin for the dwarf/giant
discrepancy.

Another possibility is that the abundances in EMP turnoff stars are
seriously affected by diffusion  (see e.g. Korn et al.  \cite{KGR}
and Lind et al.  \cite{lind}).  From Table 2 of Lind et al.
(\cite{lind}) one can deduce the following variations in abundance
ratios between TO stars and RGB stars in the globular cluster NGC
6397: $\Delta \rm [Mg/Fe] = -0.04 \pm 0.17$, $\Delta \rm [Ca/Fe] =
+0.06 \pm 0.13$, $\Delta \rm [Ti/Fe] = +0.16 \pm 0.12$.  So only for 
[Ti/Fe] is a variation marginally detected, which happens
to be of the same order of magnitude and sign as the giant/dwarf
discrepancy observed by us.

Although a role of diffusion cannot be ruled out, the evidence in
favour is, at best, very weak.  Confirmation of the results of
Korn et al.  (\cite{KGR}) and Lind et al.  (\cite{lind}) by an
independent analysis would be useful,  especially in view of the fact
that previous investigations of the same cluster (Castilho et al.  
\cite{bruno}, Gratton et al. \cite{gratton}) gave different
results.  As we pointed out in Paper VII, the adoption of a higher
effective temperature for the turn-off stars of this cluster, as done
by Bonifacio et al.  (\cite{B02}), would largely cancel the abundance
differences between TO and RGB. Even if the results for NGC 6397 were
confirmed, it is not obvious that they would apply to the field stars 
analysed in the present paper. Unlike the stars in a globular cluster, 
these stars are not necessarily strictly coeval, and their metallicities 
range from $\sim$0.7 to $\sim$1.7 dex below that of NGC 6397.  

%FIGURE 11

%\afterpage{\clearpage 
\begin {figure}
\begin {center}
\resizebox  {\hsize}{!} 
{\includegraphics[clip=true]{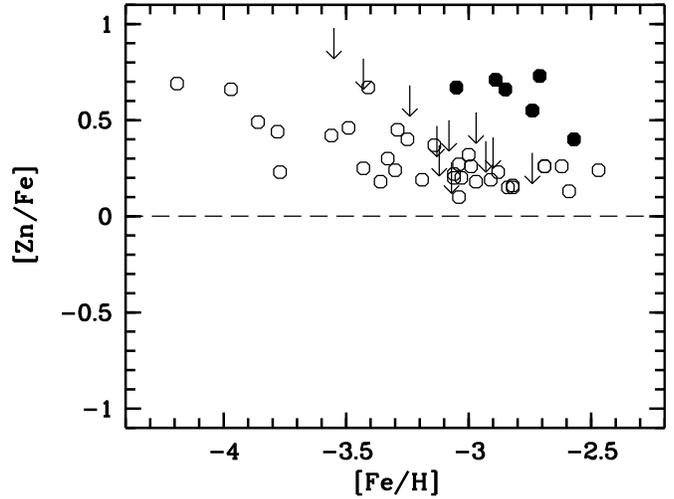} }
\caption{[Zn/Fe] ratios in dwarf stars (this paper; filled circles)
and in giants (Paper V; open circles). 
}
\label {zinc}
\end {center}
\end {figure}
%}

\subsection{Do the giant and dwarf samples belong to the same 
population?}

It could be argued that the observed giant and turnoff samples might
belong to different populations, since the giants would, on average,
be more distant than the turnoff stars.  To test this, we have
compared the radial velocities of the two samples (it would have been
preferable to compare the space velocities, but the distances and
proper motions of the giants are generally very uncertain).
Barycentric radial velocities for the turnoff stars are given in
Bonifacio et al.  (\cite{BMS06}).  For the giants they are given in
Table \ref{vr}; they are based on the yellow spectra centered at 573nm
with laboratory and measured wavelengths of numerous \ion{Fe}{i} lines
(Nave et al.  \cite{NJL94}).  The wavelengths for the telluric lines
for the zero points have been taken from Jacquinet-Husson et al.
(\cite{telluric}).  Velocity errors should be below 0.3 km~s$^{-1}$,
more than adequate for the present purpose (see also Hill et al.
\cite{HPC02}).

%FIGURE 12
\afterpage{\clearpage
\begin {figure}[t!]
\begin {center}
\resizebox  {8.0cm}{6.2cm} 
{\includegraphics {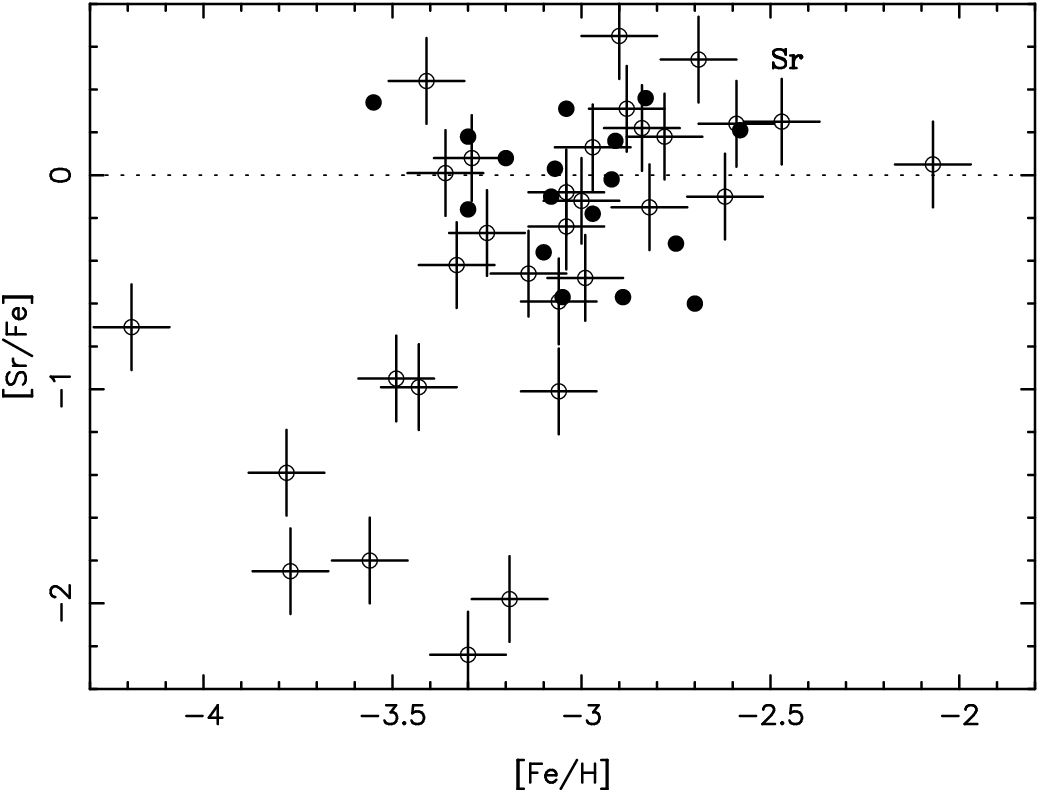} }
\resizebox  {8.0cm}{6.2cm} 
{\includegraphics {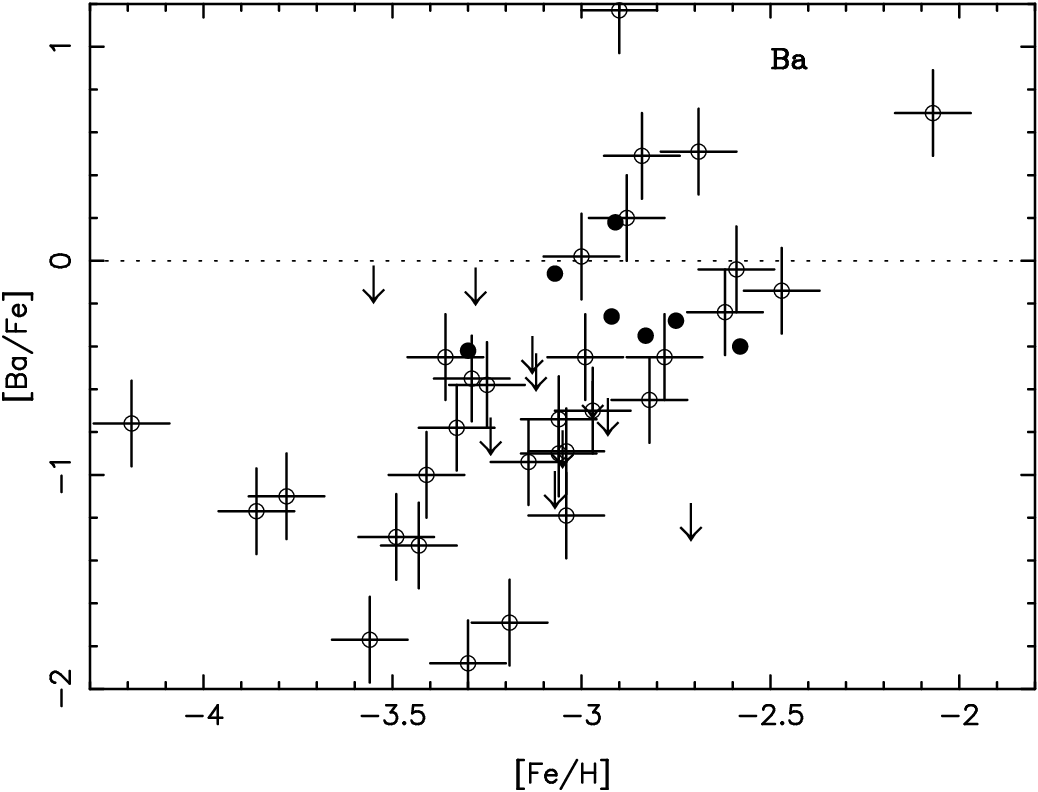} }
\caption{[Sr/Fe] and [Ba/Fe] ratios in our dwarf and giant stars; symbols 
as in Fig. \ref{carbon}. Note the large scatter: The vertical scale is 
not the 
same as in the other figures.}
\label {heavy}
\end {center}
\end {figure}
}

Since all the program stars (except for a few of the giants) have been
selected from the HK survey (Beers et al.  \cite{beers85,beers92} and 
Beers \cite{beers99}),
which is kinematically unbiased, their radial velocities should be an
unbiased estimate of the kinematic properties of the population.
Thus, if the stars were indeed drawn from different populations, we
would expect their radial-velocity distributions to differ.  The mean
radial velocities and standard deviations are -12 and 141 \kms for the
giants, -32 \kms\ and 159 \kms for the turnoff stars, respectively.  A
Kolmogorov-Smirnov test shows only a 10-15\% probability that the two
samples have not been drawn from the same parent population.  Thus,
the radial-velocity data support the assumption that the dwarfs and
giants belong to the same population.

\section{Conclusions}

We have determined abundances of C, Mg, Si, Ca, Sc, Ti, Cr, Mn, Co,
Ni, Zn, Sr and Ba for a sample of 18 EMP turnoff stars, which
complements the sample of giants discussed by Cayrel et al.
(\cite{CDS04}).  For the subgiant BS~16076-006 it was possible also to
determine the N abundance.

For Ca, Ni, Sr, and Ba we find excellent consistency between the
abundances in dwarfs and giants at any given metallicity.  For the
other elements we find abundances for the dwarfs which are about 0.2
dex larger than for giants, except for Mg and Si, for which the
abundance in dwarfs is about 0.2 dex {\em lower} than in the giants,
and Zn, for which the abundances in dwarfs are about 0.4 dex {\em
higher} than in the giants.

%FIG 13
\begin {figure}[b!]
\begin {center}
\resizebox {8.0cm}{!} 
{\includegraphics {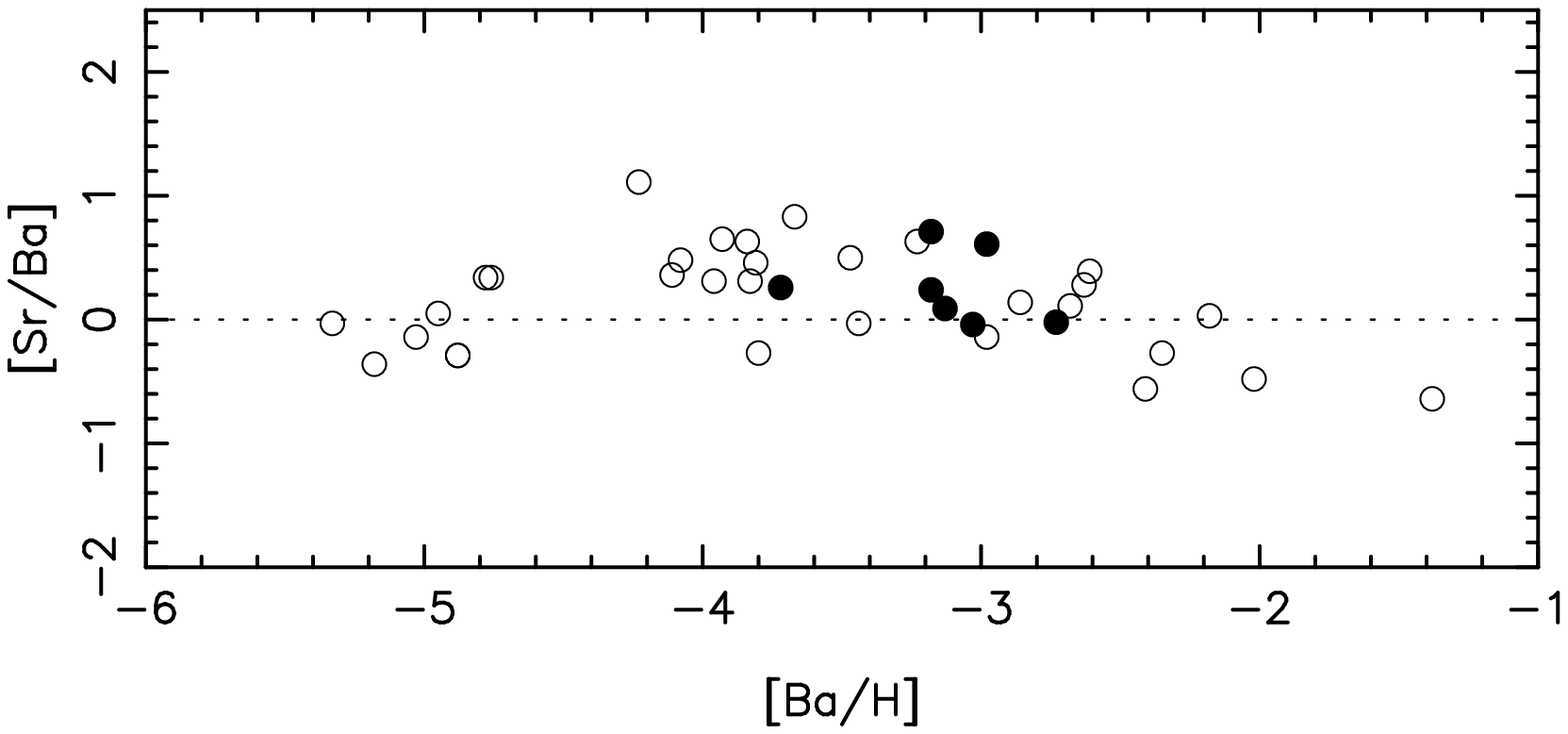} }
\caption{[Sr/Ba] as a function of [Ba/H] in our dwarf (dots) and giant 
stars (open circles; data from Paper VIII).}
\label {srba}
\end {center}
\end {figure}

The only element for which such a discrepancy could have an
astrophysical explanation is C. In fact, if the first dredge-up were
capable of bringing into the atmosphere material which had undergone
CN processing, one would expect to find lower C abundance in giants
than in dwarfs.  Such an effect is not predicted by standard models of
stellar evolution and would require some extra-mixing mechanism.  For
all the other elements which display a discrepancy between dwarfs and
giants we are unable to find any plausible astrophysical explanation.
We conclude that the discrepancies arise from shortcomings in our
analysis, probably also for C, but certainly for all other elements
for which discrepancies are found.

We have made an approximate assessment of the effects of granulation
and conclude that they are unlikely to explain the discrepancies,
except perhaps for C, Mn and Cr.  In any case, the 3D corrections
appear to be smaller for giants than for dwarfs, which suggests that
the 1D abundances of giants are preferable as reference data for
studies of the chemical evolution of the Galaxy.

The other obvious shortcoming in our analysis is the assumption of
local thermodynamic equilibrium.  Detailed NLTE line transfer
computations  exist for  Na and Al (Andrievsky et al.
\cite{ASK07,ASK08}), and for these two elements they in fact remove
the dwarf/giant discrepancy implied by the LTE analysis.
Computations for Mg are in progress, and it seems that the
agreement between giants and dwarfs is at least improved.  This
result cannot be generalized to other elements, and it is not clear
whether NLTE computations might remove any of the other discrepancies.
Clearly, NLTE computations for other key elements are urgently needed.

For readers who wish to use our data for comparison with Galactic
evolution models we suggest that, for elements for which a dwarf/giant
discrepancy exists, the abundances in giants are to be preferred.
We plan to publish an updated table of all the abundances
in the First Stars programme in a final paper
of the series. For the time being we direct the reader 
who wants the most updated  abundances of the First Stars giants,
to the following papers: for
Li, C, N and O, Spite et al. (\cite{SCP05}, First Stars VI;
\cite{SCH06}, First Stars IX); for Na,
Andrievsky et al. (\cite{ASK07}); for Mg to the NLTE abundances in 
Fig.\,\ref{MgSiCaTi} of the present paper (to be published in full 
soon); for Al, Andrievsky et al. (\cite{ASK08});
for K, Ca, Sc, Ti,  Mn, Fe, Co,  Ni, Zn,
Cayrel et al.  (\cite{CDS04}); 
for Cr, the \ion{Cr}{ii} 
abundances given in Fig.\,\ref{Cr1-2} should
be preferred; for Ba Andrievsky et al (\cite{ASK09});
for all the other elements heavier than Zn Fran\c cois et al.
(\cite{FDH06}, First Stars VIII).

  The reasons for  recommending the use of abundances in giants 
are threefold: 1) granulation effects are smaller for giants than for
dwarfs; 2) giants have lower effective temperatures and stronger
lines, so from the observational point of view their abundances are
better determined; and 3) the atmospheres of giants stars are well
mixed by convection and should be immune to chemical anomalies driven
by diffusion.  One should, however, bear in mind that future NLTE
analyses of our data could imply substantial revision of the
abundances in both giants and dwarfs.

\begin {acknowledgements} We thank the ESO staff for assistance during
all the runs of our Large Programme.  R.C., P.F., V.H., B.P., F.S. \&
M.S. thank the PNPS and the PNG for their support.  P.B., H.G.L. and
E.C. acknowledge support from EU contract MEXT-CT-2004-014265
(CIFIST).  T.C.B. acknowledges partial funding for this work from
grants AST 00-98508, AST 00-98549, AST 04-06784, AST 07-07776, as well
as from grant PHY 02-16783: Physics Frontiers Center/Joint Institute
for Nuclear Astrophysics (JINA), all from the U.S. National Science
Foundation.  B.N. and J.A. thank the Carlsberg Foundation and the
Swedish and Danish Natural Science Research Councils for partial
financial support of this work.  We acknowledge use of the
supercomputing centre CINECA, which has granted us time to compute
part of the hydrodynamical models used in this investigation, through
the INAF-CINECA agreement 2006,2007.

\end {acknowledgements}

\Online

\afterpage{\clearpage
%    TABLE 5
\begin {table}[t]
\caption {Radial velocities of the sample of EMP giant stars.}
\label {vr}
\begin {center}
\begin {tabular}{rlllrrrr}
\hline
 n &  Star       &   obs. date  &   Julian date   &    Vr   \\
   &             &              &                 & \kms\\
\hline
 1 & HD2796      &  2000-10-13  &  51830.1114692  &  -60.87 \\
 2 & HD122563    &  2000-07-18  &  51743.9645395  &  -26.39 \\
 3 & HD186478    &  2000-10-12  &  51829.9786466  &  +31.74 \\
 4 & BD+17:3248  &  2000-10-19  &  51836.9868382  & -146.55 \\
   &             &  2001-06-03  &  52063.2428859  & -146.45 \\
 5 & BD-18 5550  &  2000-10-11  &  51828.9894138  & -124.84 \\
   &             &  2001-09-05  &  52157.0417540  & -124.86 \\
 6 & CD-38 245   &  2000-07-19  &  51744.3756350  &  +46.39 \\
   &             &  2000-07-19  &  51744.3970970  &  +46.42 \\
 7 & BS16467-062 &  2001-06-04  &  52064.0495221  &  -90.58 \\
   &             &  2001-07-04  &  52094.9546310  &  -90.53 \\
 8 & BS16477-3   &  2001-06-03  &  52063.1180688  & -222.61 \\
   &             &  2001-06-04  &  52064.0946300  & -222.59 \\
 9 & BS17569-49  &  2001-06-02  &  52062.3702745  & -213.10 \\
   &             &  2001-06-03  &  52063.3972078  & -213.13 \\
10 & CS22169-35  &  2000-10-14  &  51831.2987792  &  +14.40 \\
11 & CS22172-2   &  2000-10-18  &  51835.2952562  & +251.03 \\
12 & CS22186-25  &  2001-10-19  &  52201.2882394  & -122.34 \\
   &             &  2001-10-21  &  52203.3130894  & -122.43 \\
   &             &  2001-11-07  &  52220.2859610  & -122.35 \\
13 & CS22189-09  &  2000-10-15  &  51832.2679605  &  -20.22 \\
14 & CS22873-55  &  2001-06-01  &  52061.2269695  & +214.54 \\
   &             &  2006-10-16* &  54024.9749432  & +214.23 \\
15 & CS22873-166 &  2000-10-19  &  51836.0985003  &  -17.03 \\
   &             &  2006-10-17* &  54025.9801752  &  -17.24 \\
16 & CS22878-101 &  2000-07-19  &  51744.051286   & -129.12 \\
   &             &  2000-07-19  &  51744.097628   & -129.24 \\
   &             &  2000-08-06  &  51762.984927   & -129.32 \\
17 & CS22885-96  &  2000-08-09  &  51765.0892940  & -250.28 \\
   &             &  2000-08-09  &  51765.1072672  & -250.39 \\
   &             &  2000-08-11  &  51767.0918537  & -250.01 \\
   &             &  2000-08-11  &  51767.1132987  & -249.97 \\
18 & CS22891-209 &  2000-10-18  &  51835.9948297  &  +80.29 \\
   &             &  2006-10-15* &  54023.9716681  &  +79.98 \\
19 & CS22892-52  &  2001-09-08  &  52160.1424999  &  +13.31 \\
20 & CS22896-154 &  2000-10-12  &  51829.9948968  & +137.96 \\
   &             &  2000-10-14  &  51831.9888536  & +137.82 \\
21 & CS22897-8   &  2000-10-12  &  51829.0123961  & +266.69 \\
22 & CS22948-66  &  2001-09-05  &  52157.1005735  & -170.67 \\
   &             &  2006-10-18* &  54026.1255044  & -170.62 \\
23 & CS22949-37  &   \multicolumn{2}{c}{(see Depagne et al., 2002)}& 
-125.64 \\
24 & CS22952-15  &  2000-10-14  &  51831.0576878  &  -18.24 \\
   &             &  2006-10-18* &  54026.1033137  &  -18.43 \\
25 & CS22953-3   &  2001-09-08  &  52160.3271107  & +208.61 \\
   &             &  2001-09-09  &  52161.3657380  & +208.44 \\
26 & CS22956-50  &  2001-09-04  &  52156.1279571  &   -0.09 \\
27 & CS22966-57  &  2001-09-04  &  52156.1945760  & +101.22 \\
28 & CS22968-14  &  2000-10-12  &  51829.3235079  & +159.16 \\
   &             &  2000-10-12  &  51829.3662157  & +159.09 \\
   &             &  2000-10-17  &  51834.2971840  & +159.16 \\
   &             &  2001-11-09  &  52222.2098787  & +159.02 \\
29 & CS29491-53  &  2001-10-20  &  52202.0858094  & -147.47 \\
30 & CS29495-41  &  2001-06-02  &  52062.2774334  &  +79.61 \\
31 & CS29502-42  &  2000-10-12  &  51829.0798142  & -138.05 \\
   &             &  2001-09-06  &  52158.1633788  & -138.16 \\
   &             &  2001-09-09  &  52161.1907037  & -138.07 \\
32 & CS29516-24  &  2001-06-04  &  52064.3718149  &  -84.27 \\
33 & CS29518-51  &  2000-10-13  &  51830.1969033  &  +96.42 \\
34 & CS30325-94  &  2000-08-08  &  51764.9714377  & -157.76 \\
   &             &  2000-08-08  &  51764.9928808  & -157.67 \\
35 & CS31082-01  &  \multicolumn{2}{c}{(see Hill et al., 2001)}& 
+139.07 \\
\hline
\end {tabular}
\end {center}
\end {table} 

%*****************************************************************
}

\onecolumn

\appendix
\section{Comparison with the 0Z project. \label{0Zcomp}}

\begin {figure}
\begin {center}
\resizebox{\hsize}{!}
{\includegraphics[clip=true]{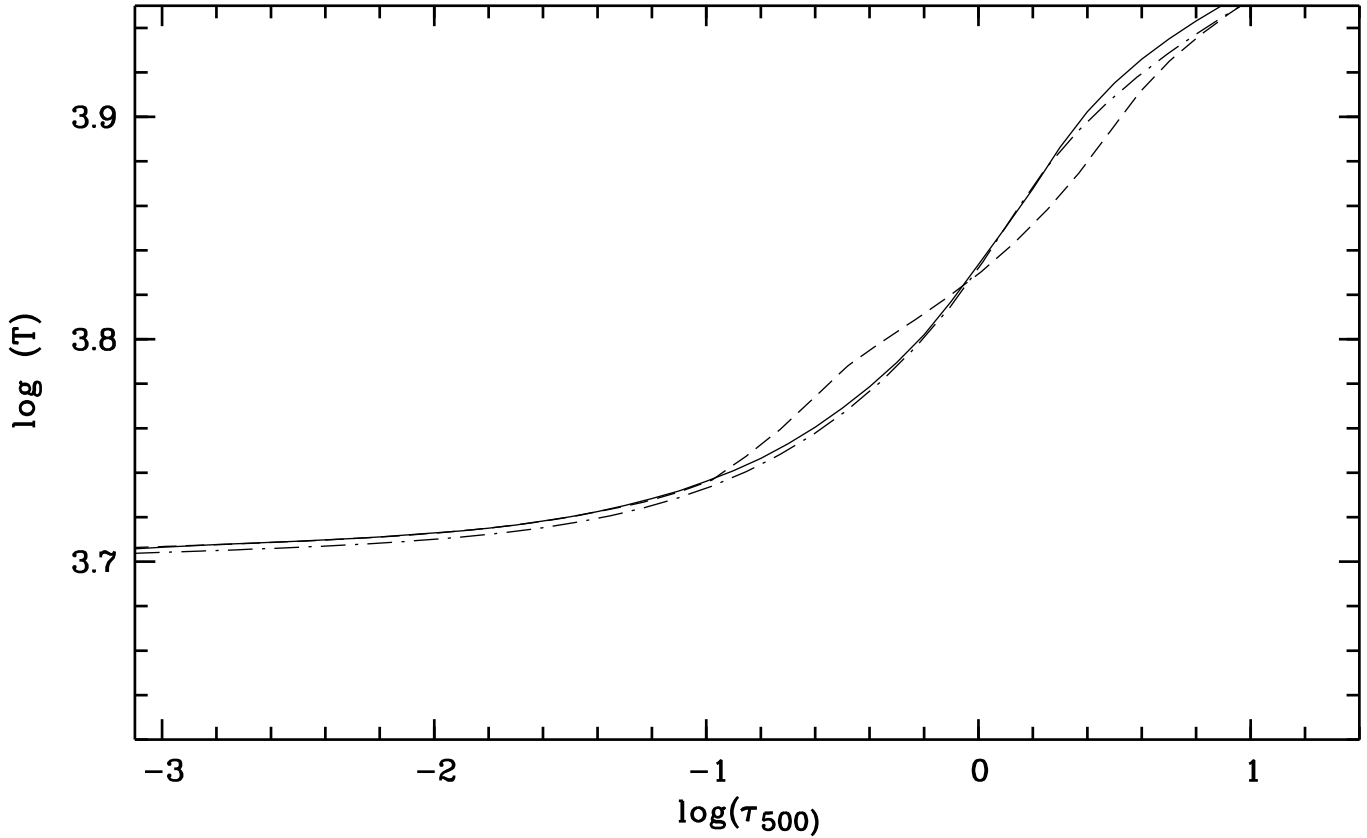}}
\caption{Temperature structure for three models with 
\teff = 6365, \logg 4.4 and [M/H]= --3.0.
The solid line is our MARCS models, the dashed line
is an atlas overshooting model, the dashed-dotted line
is an ATLAS non-overshooting model.}
\label {over_dwarfs}
\end {center}
\end {figure}
\begin {figure}
\begin {center}
\resizebox{\hsize}{!}{\includegraphics[clip=true]{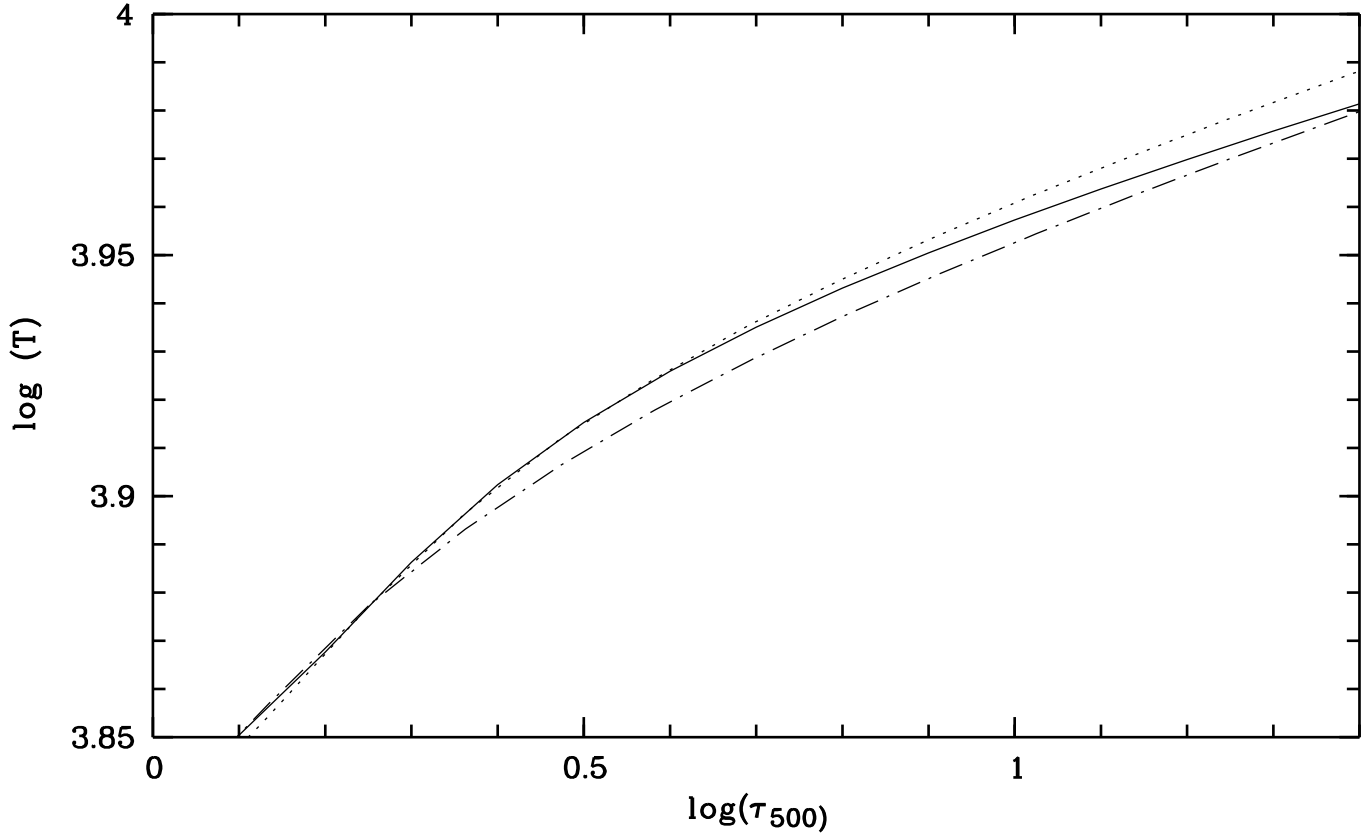}}
\caption{Temperature structure of the
deepest layres of three models with 
\teff = 6365, \logg =  4.4 and [M/H]= --3.0.
The solid line is our MARCS models, the dashed-dotted line
is an ATLAS non-overshooting model with \mlp = 1.25 (also
shown in Fig. \ref{over_dwarfs}), the dotted line
is an ATLAS non-overshooting model with \mlp = 1.00.}
\label {ML}
\end {center}
\end {figure}

The 0Z project (Cohen et al.  \cite{CCM04,CCM08}) has produced a data
set similar to that of the ``First Stars'' project, it is therefore of
some interest to verify how these data sets compare.  Cohen et al.
(\cite{CCM04}) analysed a set of dwarf stars which is directly
comparable to those analysed in the present paper.  The spectra were
acquired with the HIRES spectrograph at the Keck I Telescope, at a 
resolution only slightly lower than our UVES-VLT data (34\,000
rather than 45\,000), and the S/N ratios are comparable.  The
equivalent widths were measured using an automatic code which fits
gaussians, therefore the general philosphy of EW measurement does not
differ from ours.  In fact Cohen et al.  (\cite{CCM08}) observed
two giant stars measured by Cayrel et al.  (\cite{CDS04}), and the
equivalent widths compare very well (see figures 13 and 14 of Cohen et
al.  \cite{CCM08}, and related text).  The two projects differ in the
method used to fix the atmospheric parameters: we use the wings of
H$\alpha$ for dwarf stars, while the 0Z project relies on photometry
to derive \teff.  For surface gravity we use the iron ionisation
equilibrium, while the 0Z project relies on theoretical isochrones.

We have also investigated the $gf$ values used by the two projects, and
they are very similar; the use of the one or the other set would not
imply differences in the derived abundances smaller or equal to
0.02\,dex.

Thus, part of the differences will depend on the different adopted
atmospheric parameters.  There is no dwarf star in common between the
two groups; thus it is not straightforward to compare the results of
the two projects.

For the analysis the two projects use different model atmospheres and
different line formation codes.  We use MARCS model atmospheres and
{\tt turbospectrum}, while the 0Z project uses ATLAS models
interpolated in the grid of Kurucz (1993), with the overshooting
option switched on, and the MOOG code (Sneden
\cite{Sneden73,Sneden74,snedenweb}).  As we shall show below, the
different choice of line formation is relatively unimportant, implying
differences in the abundances of a few hundredths of dex; on the other
hand the choice of ATLAS overshooting models implies abundances which
are higher by about 0.1\,dex for all the models.  Such a behaviour was
already noticed by Molaro et al.  (\cite{molaro}) for Li, but we show
here that it is indeed true for all species.

\begin {table}[t]
\caption {Abundances for HE 0508-1555 for different model atmospheres.}
\label {HE0508}
\centering
\begin {tabular}{lllllllcc}
\hline
\hline
Ion & A &$\rm \sigma_A$ &A &$\rm \sigma_A$ &A &$\rm \sigma_A$ &A &$\rm \sigma_A$  
\\
    & \multicolumn{2}{c}{MARCS} 
&\multicolumn{2}{c}{ATLAS}&\multicolumn{2}{c}{ATLAS}&\multicolumn{2}{c}{Cohen}\\
    &          &                 
&\multicolumn{2}{c}{NOVER}&\multicolumn{2}{c}{OVER}&\multicolumn{2}{c}{et al. 
2004}\\ 
\hline
%         gf cohen                       Cohen 2004    
%       Ab      err                 
\ion{Mg}{i}  &5.58&  0.27 & 5.54&0.29&5.68&0.27 & 5.70  0.28\\
\ion{Al}{i}  &3.28&  0.06 & 3.24&0.06&3.38&0.06 & 3.91  0.09\\
\ion{Si}{i}  &5.19&       & 5.14&    &5.31&     & 5.23      \\
\ion{Ca}{i}  &4.15&  0.08 & 4.11&0.08&4.24&0.08 & 4.22  0.11\\
\ion{Sc}{ii} &0.99&  0.03 & 0.96&0.04&1.04&     & 1.02  0.03\\
\ion{Ti}{i}  &3.18&  0.13 & 3.14&0.13&3.26&0.13 & 3.23  0.14\\
\ion{Ti}{ii} &3.05&  0.09 & 3.02&0.09&3.10&0.09 & 3.08  0.10\\
\ion{Cr}{i}  &2.94&  0.03 & 2.89&0.03&3.01&0.03 & 2.99  0.04\\
\ion{Mn}{i}  &2.31&  0.05 & 2.26&0.05&2.38&0.04 & 2.35  0.05\\
\ion{Fe}{i}  &5.00&  0.16 & 4.96&0.16&5.09&0.15 & 5.07  0.17\\
\ion{Fe}{ii} &5.11&  0.16 & 5.09&0.16&5.16&0.16 & 5.13  0.17\\
\ion{Co}{i}  &2.88&  0.11 & 2.84&0.11&2.96&0.10 & 2.92  0.11\\
\ion{Ni}{i}  &3.77&       & 3.72&    &3.86&     & 3.83      \\   
\hline
\end{tabular}
\end{table}

In Table \ref{HE0508} we list the abundances for the star
HE\,0508-1555 derived by using the equivalent widths of Cohen et al.
(\cite{CCM04}) and their atmospheric parameters (\teff = 6365, log g =
4.4 and a microturbulent velocity of 1.6 \kms) with three different
models: a MARCS model interpolated in our grid, an ATLAS model
computed without overshooting and an ATLAS model computed with
overshooting.  For all the models we assumed [M/H]=--3.0.  Our ATLAS
models are somewhat different from those of the Kurucz (1993) used by
the 0Z project.  In the first place we use the ``NEW'' opacity
distribution functions (Castelli \& Kurucz \cite{CK}) computed with 1
\kms\ microturbulence.  In the second place we use the Linux version
of ATLAS (Sbordone et al.  \cite{SBCK}).  In all cases the line
formation code used was {\tt turbospectrum}.  In the last two columns
of Table \ref{HE0508} we provide the abundances of Cohen et al.
(\cite{CCM04}), for the reader's convenience.  

Inspection of Table
\ref{HE0508} immediately suggests that both the difference in ATLAS
versions and the different line formation codes used are immaterial,
since the abundances we find for almost all elements are within 0.04
dex of those of Cohen et al.  (\cite{CCM04}).  The two exceptions are
Al and Si.  For Al there is a good reason for the discrepancy: Both
\ion{Al}{i} lines used are affected by the neighbouring Balmer
lines.  In our analysis we used spectrum synthesis to derive the
abundances.  Instead, MOOG can take into account the absorption due to
the Balmer lines, either using the {\bf opacit} switch to introduce a
fudge factor on the continuum opacity or using the {\bf strong}
keyword to read strong lines to be considered.  

For Si the difference
between our result with the ATLAS overshooting model and the
published value of Cohen et al.  (\cite{CCM04}) is 0.08 dex.  This
abundance is based on a single line of about 10\,pm of EW, therefore
clearly saturated.  The precise value of the damping constants used
for this line and the way the different codes use them may have an
impact.  

Another inference which can be drawn from Table \ref{HE0508}
is that MARCS models and ATLAS non-overshooting models provide results
which are quite similar.  This is not the case for the ATLAS
overshooting models, which imply abundances which are higher by about
0.1\, dex for all elements.  The reason for this behaviour may be
understood by looking at the temperature structure of the different
models.  In Fig.  \ref{over_dwarfs} we compare the temperature
structures of our MARCS model (solid line), the ATLAS non-overshooting
model (dashed-dotted line) and the ATLAS overshooting model (dashed
line).  The temperature structure of the ATLAS non-overshooting and of
the MARCS model are quite similar.  In fact, the only difference is for
the deepest layers and is driven by the different choice made for the
mixing length.  

In Fig.  \ref{ML} we show the temperature structure of
the deeper layers of our MARCS model (solid line) together with two
ATLAS non-overshooting models with different values of \mlp\, 1.25
(dashed-dotted line) and 1.00 (dotted line).  The ATLAS model with
\mlp = 1.00 is closer to the MARCS model, up to $\log\tau_{500}\sim
0.7$, but then becomes hotter than the MARCS model.  In general it is
impossible to chose a \mlp\ such as a MARCS and an ATLAS model have
exactly the same structure in depth, due to the different formulation
of the mixing-length therory in the two codes.  Such differences in
the very deepest layers have very little influence on a typical
abundance analysis.  In fact only the lines which form in these very
deep layers are affected, i.e. very weak lines of 0.1\,pm or smaller,
and the wings of H$\beta$ and higher members of the Balmer series.

In general we can conclude that MARCS and ATLAS non-overshooting
models are very similar, and an abundance analysis based on the two
models will yield abundances which are consistent within a few
hundredths of dex.  The situation is dramatically different when we
consider the ATLAS overshooting models.  Such models present a
temperature structure which is very different from both ATLAS
non-overshooting and MARCS models in the region $-1 \le \log\tau_{500}
\le 1$ where the majority of lines used in abundance analysis are
formed.  

Castelli, Gratton \& Kurucz (\cite{CGK,CGKE}) 
investigated extensively the effects of the approximate overshooting
present in ATLAS and concluded that the no-overshooting models are
capable of reproducing a larger set of observables, thus discouraging
the use of overshooting models.  To these considerations we may add
that having investigated the mean temperature structures of \cobold\
3D hydrodynamical models we never see the typical ``bump'' in the
temperature structure seen in ATLAS overshooting models.  The real
effect of the overshooting is the over-cooling of the outer layers
with respect to what is predicted in radiative equilibrium models
(Asplund et al.  \cite{asp99}, Collet et al.  \cite{collet}, Caffau \&
Ludwig \cite{CL07}, Gonz\'alez Hern\'andez et al.  \cite{jonay}, Paper
XI).  This is a further reason to avoid the use of the ATLAS
overshooting models.

It can be appreciated that the differences due to different models
largely cancel out when considering abundance ratios, like e.g.
[Mg/Fe], rather than abundances relative to hydrogen.  For example
[Mg/H] is -2.04 for the ATLAS non-overshooting model, but -1.90 for
the ATLAS overshooting one, however [Mg/Fe] is 0.50 \relax in the
first case and 0.51 \relax in the second case.

A difference in the average [Mg/Fe] is found between us and the 0Z
project, of the order of 0.2\,dex (the 0Z project being higher), both
if we consider only dwarf stars, only giants, or the full samples.
Such an offset is roughly compatible with a $1\sigma$ error on each
side, but perhaps a little disturbing.  Only a 0.01\, dex difference
is due to the different adopted solar abundances.  The use of
different models and different atmospheric parameters should largely
cancel out when considering a ratio such as [Mg/Fe].  Largely does not
mean totally, however: Table \ref{BS} shows a 0.06\, dex
difference in [Mg/Fe] for BS 16467-062, depending on the adopted
atmospheric parameters.  

Table 10 of Cohen et al.  (2004) is also
illuminating; it shows how the average [Mg/Fe] changes if one
considers the mean computed from the abundances derived from a single
line of \ion{Mg}{i}.  Of the five \ion{Mg}{i} lines used by Cohen et
al.  (2004) three tend to give systematically higher abundances, while
two give systematically lower abundances.  The final result depends on
the set of adopted lines.  This issue requires further investigation
in the light of the study of deviations from thermodynamic equilibrium
for the \ion{Mg}{i} lines.  Our abundance ratios are in agreement with
those provided by the 0Z project, within the stated errors.

At the end of this exercise we conclude that our measurements and
those of the 0Z Team are highly consistent.  Differences in the
published abundances can be traced back to the different atmospheric
parameters adopted, the different treatment of convection in the
adopted model atmospheres (approximate overshooting versus no
overshooting), and for some elements to the particular choice of lines.

\section{Details of the comparison with Lai et al. 2008\label{laicomp}}

\begin {figure}
\begin {center}
\resizebox{\hsize}{!}{\includegraphics[clip=true]{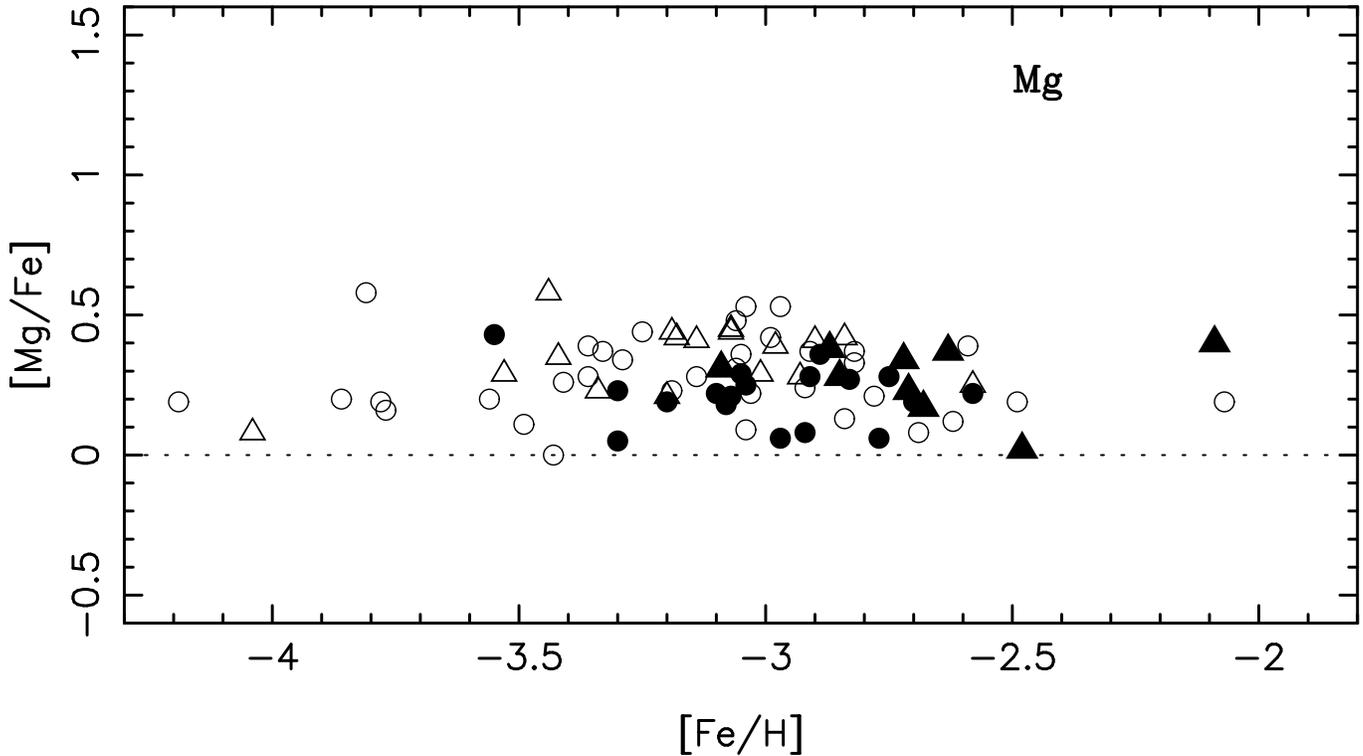}}
\caption{Comparison of the [Mg/Fe] ratios of the First Stars project
and those of Lai et al. (2008). Our data is shown as circles, 
while those of Lai et al. as triangles. Open symbols
correspond to giant stars, while filled symbols to dwarfs.}
\label {LaiMg}
\end {center}
\end {figure}

\begin {figure} \begin {center}
\resizebox{\hsize}{!}{\includegraphics[clip=true]{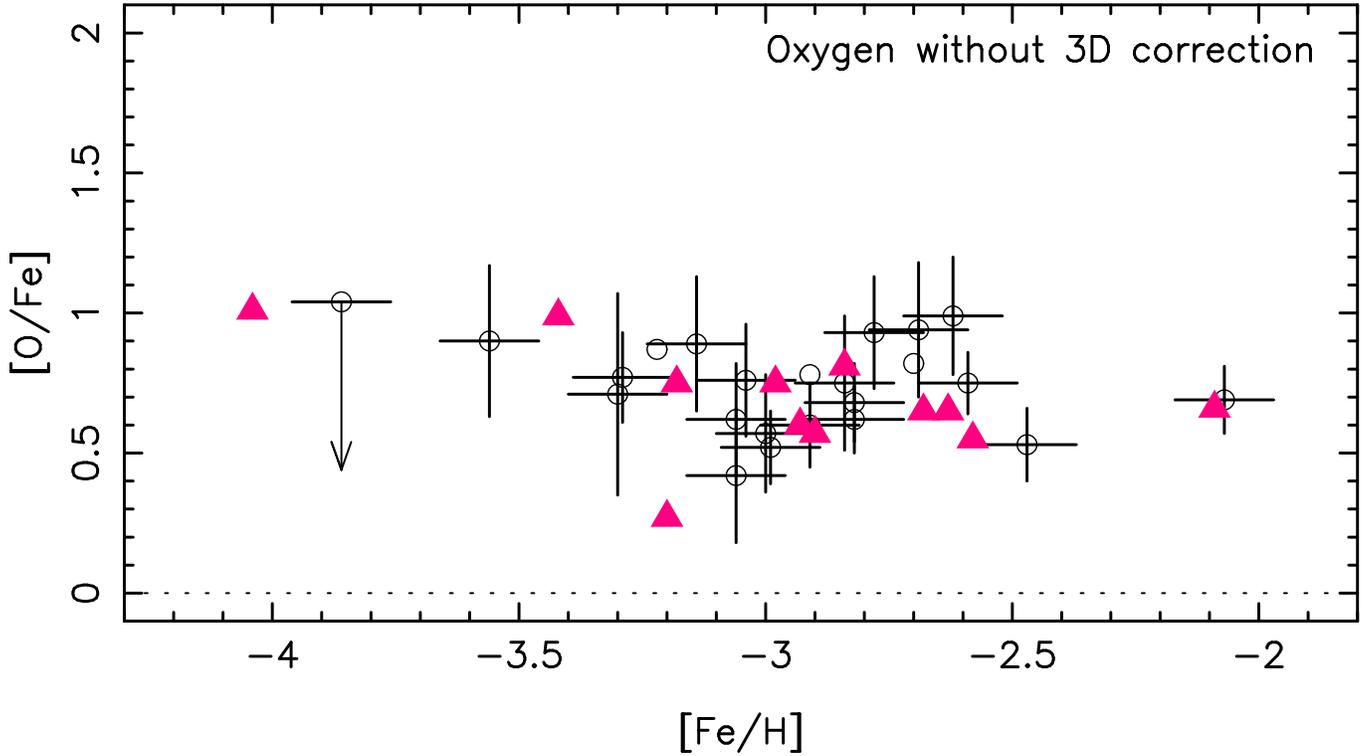}}
\caption{Comparison of the [O/Fe] ratios of the First Stars project
and those of Lai et al.  (2008).  Symbols as in Fig.  \ref{LaiMg}.
Our oxygen abundances are derived from the 630\,nm [OI] line, while
those of Lai et al.  from one UV OH line of the $A^2\Sigma - X^2\Pi$
electronic system around 318.5\,nm.} 
\label {LaiO} 
\end {center} 
\end{figure}

\begin {figure*}
\begin {center}
\resizebox{\hsize}{!}{\includegraphics[clip=true]{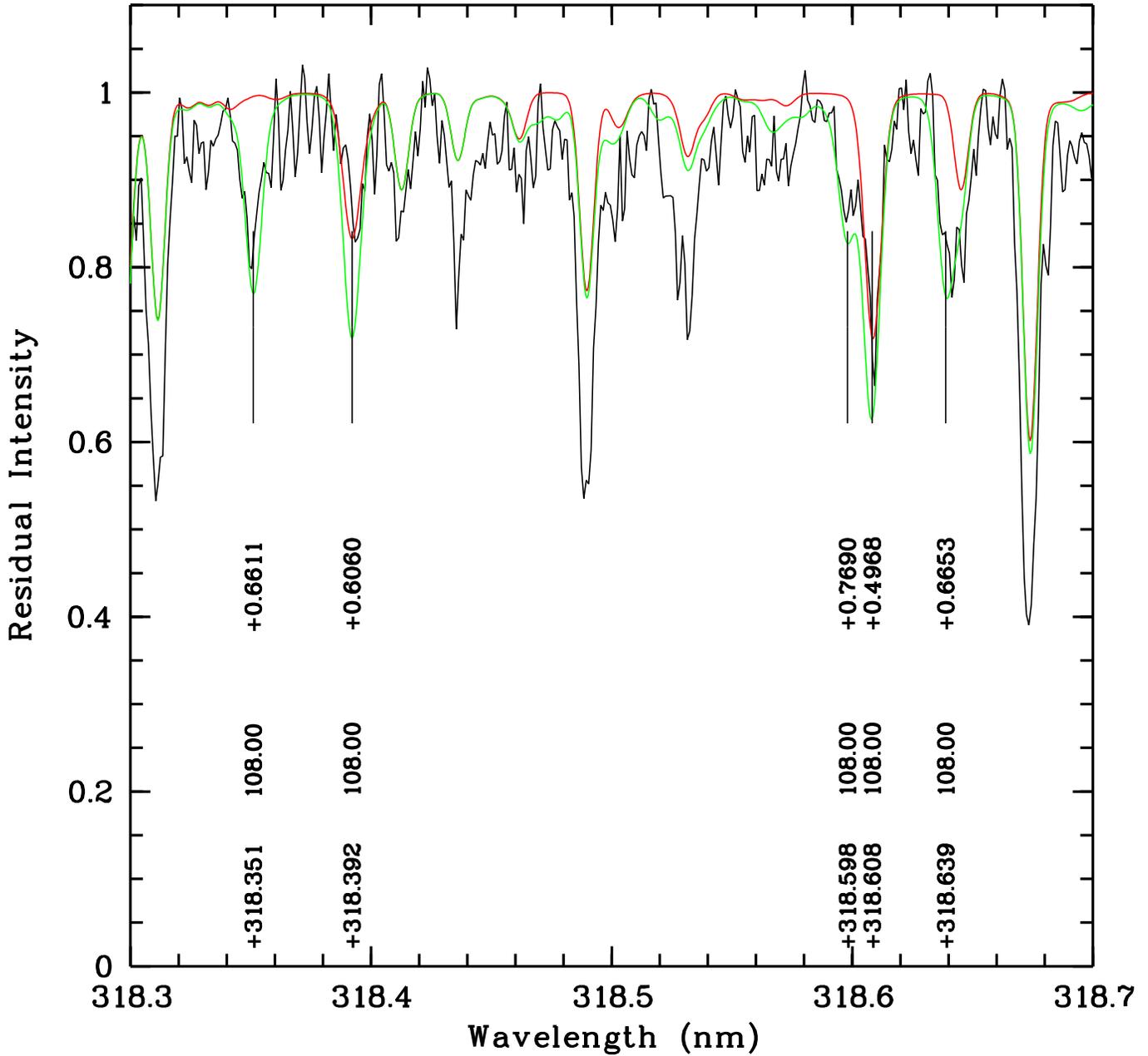}}
\caption{HIRES-Keck spectrum of the dwarf star CS 31085-024.  The data
is the same used by Lai et al.  (2008), downloaded from the Keck
Observatory Archive (http://www2.keck.hawaii.edu/koa/public/koa.php).
Overimposed on the spectrum are two synthetic spectra, computed with
SYNTHE, from an ATLAS 9 model with \teff = 5949, \logg = 4.57 and
metallicity --3.0 and [O/Fe]=1.54.  The synthetic spectrum plotted in
red has been computed using the $gf$ values for the OH lines of the
(0-0) vibrational band of the $A^2\Sigma - X^2\Pi$ electronic system
computed from the lifetimes of Goldman \& Gillis (\cite{GG}).  Instead
the one in green has been computed using the line list of the OH
$A^2\Sigma - X^2\Pi$ computed by R.L. Kurucz and distributed through
(http://wwwuser.oats.inaf.it/atmos/tarballs/molecules.tar.bz2).
}
\label{CS31085-024}
\end {center}
\end {figure*}

\begin {figure}
\begin {center}
\resizebox{\hsize}{!}{\includegraphics[clip=true]{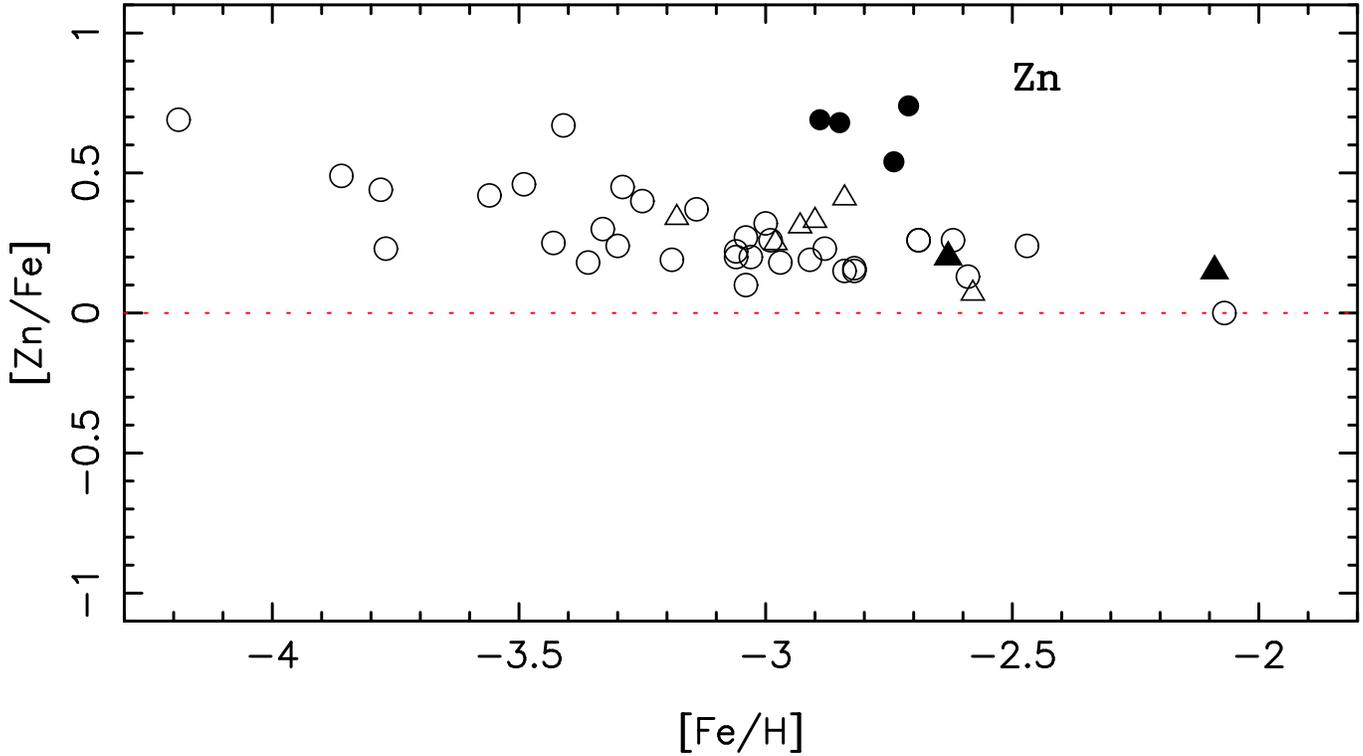}}
\caption{Comparison of the [Zn/Fe] ratios of the First Stars project
and those of Lai et al. (2008). Symbols as in Fig. \ref{LaiMg}.
}
\label {LaiZn}
\end {center}
\end {figure}

Lai et al.  (2008) also analysed a set of stars which is comparable to
that of the First Stars project with respect to metallicity.
Their sample is also extracted from the HK survey and comprises both
dwarfs and giants.  Their method to determine atmospheric parameters
is similar to that of the 0Z project, photometric temperatures from
the $V-K$ colour and gravities derived from isochrones.  They observe
the giant star BS 16467-062, also observed by us (Paper V) and in the 
0Z project (Cohen et al.  2008) and, not surprisingly, derive
atmospheric parameters very close to those of Cohen et al.  This
allows a very tight comparison of the analysis by the three groups,
which we defer to Sec.  \ref{BScomp}.

Lai et al.  use the same spectrum synthesis code as we and also use
ATLAS 9 non-overshooting models which, as discussed in Sec.
\ref{0Zcomp}, are very similar to our MARCS models.  It is therefore
to be expected that the abundance ratios determined by the two groups
are quite similar.  In Fig.  \ref{LaiMg} we compare the [Mg/Fe] ratios
of the First Stars project with those of Lai et al.  The overall
agreement is satisfactory.

In Fig.  \ref{LaiO} we compare the [O/Fe] ratios of the First Stars
project (only giants) with those of Lai et al.  The figure seems to
indicate a good agreement; however we believe that this agreement is in
fact fortuitous, as  
our oxygen abundances were based on the
630\,nm [OI] line, while those of Lai et al.  have been derived from
one OH line of the UV $A^2\Sigma - X^2\Pi$ electronic system around
318.5\,nm (although the precise line used is not specified).  These OH
lines are known to provide very high [O/Fe] ratios when analysed with
1D model atmospheres (e.g. Boesgaard et al.  \cite{boe99}, Israelian
et al.  \cite{isr01}).  Asplund \& Garc{\'\i}a P{\'e}rez
(\cite{agp01}) have explained this behaviour as due to overcooling
of the outer layers of the stars, caused by the
overshooting of the convective elements and not properly described by
1D model atmospheres.  Our own hdyrodynamical computations (Gonz\'alez
Hern\'andez et al.  \cite{jonay}, Paper XI) confirm this
interpretation.  In view of this fact it is, at first sight,
surprising to find that Lai et al.  determine rather low [O/Fe]
ratios from the OH lines.  Closer inspection of their analysis reveals,
however, that this is mainly driven by their adopted $gf$ values for
these lines.

In Fig.  \ref{CS31085-024} we show a portion of the spectrum of CS
31085-024, used by Lai et al., which we downloaded from the Keck
Observatory Archive
\footnote{http://www2.keck.hawaii.edu/koa/public/koa.php} compared
with two synthetic spectra computed using an ATLAS 9 model with the
atmospheric parameters adopted by Lai et al.  and two different OH
line lists.  In the first case (red line) we adopted the $gf$ values
for the OH lines of the (0-0) vibrational band of the $A^2\Sigma -
X^2\Pi$ electronic system computed from the lifetimes of Goldman \&
Gillis (\cite{GG}), which we used in Paper IX. In the second case
(green spectrum) we used the lines computed by R.L. Kurucz.  This
second list is far richer, since it includes also lines from other
vibrational bands and not only the (0-0) band. However, even from this
limited portion of the spectrum it can be appreciated that the Kurucz
$gf$ values are larger than those derived from the Goldman \& Gillis
(\cite{GG}) lifetimes;  use of the latter $gf$ values would lead to
considerably larger OH abundances.  

For this reason we believe that
the oxygen abundances in the stars of the Lai et al.  sample should be
reinvestigated using a different set of $gf$ values and hydrodynamical 
model atmospheres.  It is likely that the 3D corrections for
the giant stars (the majority of the Lai et al.  sample with oxygen
measurements) are smaller than those for dwarf stars (see Paper XI),
since the overcooling is far less extreme in giants than in dwarfs, it
is however unlikely that the effect is negligible.  

We disagree with
the statement by Lai et al., who discard the use of 3D models for the
analysis of the OH lines since ``these models seem to overpredict the
solar oxygen abundance derived from helioseismology (Delahaye \&
Pinsonneault 2006)''.  In the first place the oxygen abundance in the
Sun is not derived from OH UV lines; in the second place it is now
clear that the low solar oxygen abundances which have been claimed in
the past (Asplund et al.  2004) are not due to the use of 3D
hydrodynamical models, but to low measured EWs and extreme assumptions
on the role of collisions with H atoms in the NLTE computations (see
Caffau et al.  2008, for a discussion and a new measurement of the
solar oxygen abundance).  In our view the use of 3D hydrodynamical
models is necessary for a reliable analysis of OH lines in metal-poor
stars.

The [Cr/Fe] ratios were compared in Fig.  \ref{CrLai} and we see that
the picture which emerges is very consistent between the two analyses,
including the dwarf--giant discrepancy discussed in Sec.
\ref{CrCoNi}.  In agreement with us, Lai et al.  note that when
\ion{Cr}{ii} lines are measurable the [\ion{Cr}{ii}/Fe] ratio remains
close to zero, suggesting that the decrease in [Cr/Fe] with decreasing
metallicity, seen when \ion{Cr}{i} lines are used, is probably an
artifact due to deviations from LTE.

Finally in Fig.  \ref{LaiZn} we compare the [Zn/Fe] ratios with those
of Lai et al.  (2008).  The have measured Zn only in two dwarfs,
slightly more metal-rich than our ones and the Zn abundances for these
two are in line with what derived from the giants.  We note that the
$gf$ value adopted by Lai et al.  is 0.04\,dex lower than adopted
by us.

\section{Comparison for BS 16467-062 \label{BScomp}}

\begin {table}[t]
\caption {Abundances for BS 16467-062 different model atmospheres and the
EWs of Cohen et al. (2008).}
\label {BS}
\centering
\begin {tabular}{lrrr}
\hline
\hline
Ion & MARCS & ATLAS & MARCS\\
    &       &  OVER &      \\
    & T=5364K&T=5364 & T=5200 \\
    & log g =2.95 & log g =2.95 & log g = 2.50\\ 
\hline
\ion{Mg}{i} &  4.24  &  4.39  & 4.15\\
\ion{Al}{i} &  2.01  &  2.15  & 1.85\\
\ion{Si}{i} &  4.20  &  4.38  & 4.07\\
\ion{Ca}{i} &  2.83  &  2.98  & 2.71\\
\ion{Sc}{ii}& -0.29  & -0.20  &-0.54\\ 
\ion{Ti}{i} &  1.71  &  1.83  & 1.52\\
\ion{Ti}{ii}&  1.66  &  1.74  & 1.42\\
\ion{Cr}{i} &  1.55  &  1.68  & 1.37\\
\ion{Mn}{i} &  1.18  &  1.29  & 0.97\\
\ion{Fe}{i} &  3.87  &  4.00  & 3.70\\
\ion{Fe}{ii}&  3.93  &  4.02  & 3.73\\
\ion{Co}{i} &  2.06  &  2.20  & 1.86\\
\hline
\end{tabular}
\end{table}
\begin {figure}
\begin {center}
\resizebox{\hsize}{!}{\includegraphics[clip=true]{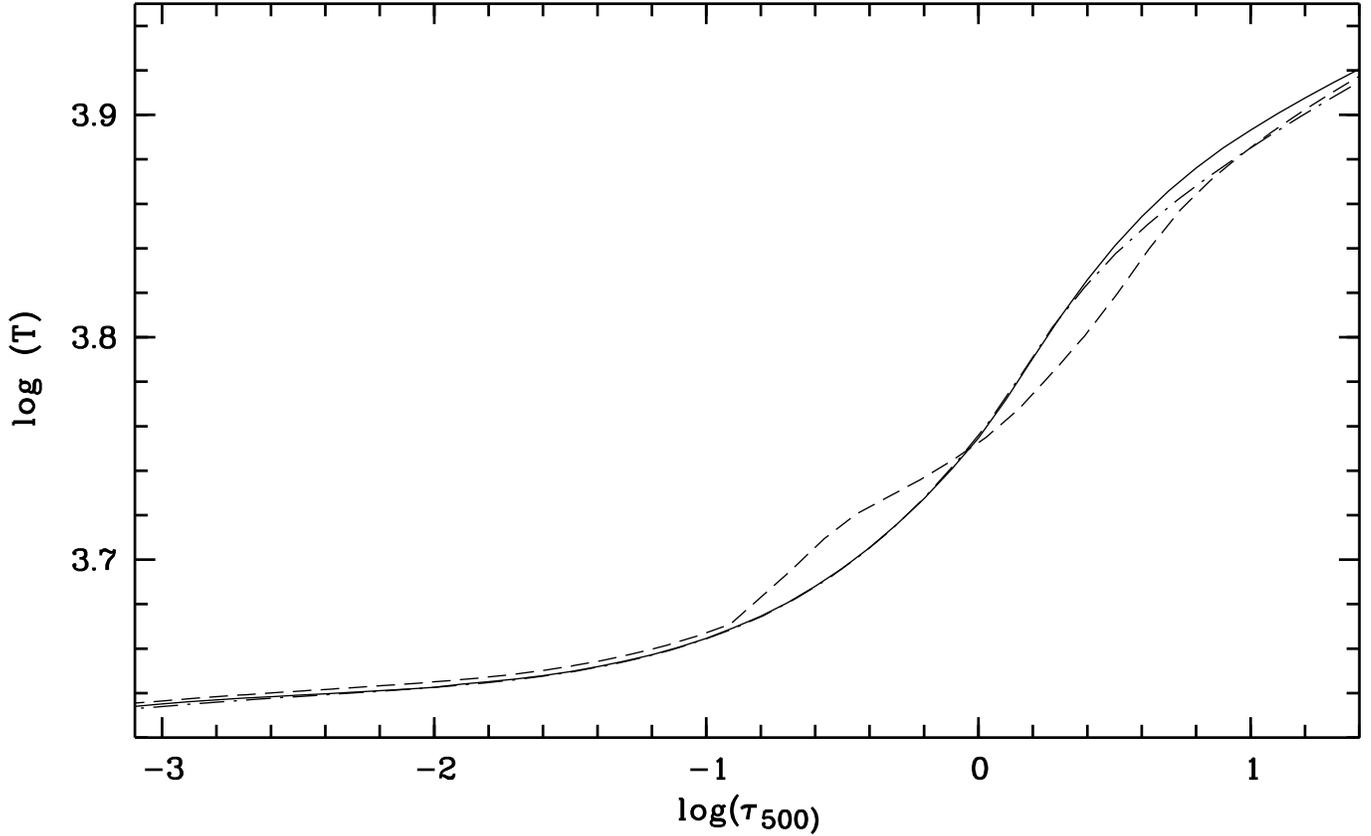}}
\caption{Temperature structure for three models with 
\teff = 5365, \logg 2.95 for BS 16467-062.
The solid line is our MARCS models, the dashed line
is an ATLAS overshooting model, the dashed-dotted line
is an ATLAS non-overshooting model.}
\label {over_giants}
\end {center}
\end {figure}

\begin {table}[t]
\caption {Abundances for BS 16467-062 for different atmospheric
parameters and the EWs of Lai et al. (2008).}
\label {BS_2}
\centering
\begin {tabular}{lrrrr}
\hline
\hline
Ion & \multicolumn{2}{c}{ATLAS} & \multicolumn{2}{c}{MARCS}\\
    & \multicolumn{2}{c}{NOVER} &      \\
    & \multicolumn{2}{c}{T=5388} & \multicolumn{2}{c}{T=5200} \\
    & \multicolumn{2}{c}{log g =3.04} &  \multicolumn{2}{c}{log g = 2.50}\\ 
             &    A & $\sigma$ & A & $\sigma$\\

\hline
\ion{Mg}{i}  &   4.06&0.10  &  3.98&0.10\\  
\ion{Si}{i}  &   4.12&      &  4.01     \\
\ion{Ca}{i}  &   2.90&0.09  &  2.78&0.07\\
\ion{Sc}{ii} &  -0.30&0.02  & -0.59&0.02\\ 
\ion{Ti}{i}  &   1.79&      &  1.57     \\
\ion{Ti}{ii} &   1.71&0.12  &  1.43&0.12\\
\ion{Cr}{ii} &   1.47&0.06  &  1.27&0.07\\
\ion{Mn}{i}  &   1.05&0.01  &  0.82&0.02\\
\ion{Fe}{i}  &   3.77&0.19  &  3.61&0.18\\ 
\ion{Fe}{ii} &   3.76&0.15  &  3.54&0.16\\ 
\ion{Co}{i}  &   1.88&0.06  &  1.66&0.06\\
\ion{Ni}{i}  &   2.80&0.02  &  2.60&0.04\\
\hline
\end{tabular}
\end{table}

The giant star BS 16467-062 has been 
observed independently by all three groups, 
ourselves (Cayrel et al. 2004, Paper V),
the 0Z project (Cohen et al. 2008) and
Lai et al. (2008). The latter two groups
have used HIRES@Keck, while we have used
UVES@VLT.

In their appendix B, Cohen et al. (\cite{CCM08}) 
make a detailed comparison between the analysis of giant
stars analysed by us and their own analysis.
They conclude that the same star analysed by
the two groups will show a difference of 0.3\,dex
in [Fe/H]. This is based on their analysis
of the giant star BS 16467-062. We wish to explain
how this difference arises. 
We used the EWs of Cohen et al. (\cite{CCM08}) for
this star and their gf values to redetermine
the abundances using four models: 
a MARCS model and two ATLAS models (overshooting
and non-overshooting) for \teff = 5365 K \logg = 2.95,
which are the parameters of  Cohen et al. (\cite{CCM08})
and the MARCS model with \teff = 5200 K , \logg = 2.50,
which was used in Cayrel et al. (\cite{CDS04}, Paper V).
The results are shown in Table \ref{BS}.
We omit the results from the ATLAS non-overshooting model,
since they are identical to those obtained
from the MARCS model. 
This could be expected by looking at Fig. \ref{over_giants}
in which the temperature structures of the two models are 
compared.

The differences in the abundances between the MARCS
model with the parameters of Paper V and those
from an ATLAS overshooting model with the higher
\teff\ and \logg\ of Cohen et al. (\cite{CCM08})
may indeed be as large as 0.3\,dex. 
However, it is important to understand that this difference 
is due to two distinct factors: on the one hand
the change in \teff\ and \logg, as each species displays
a slightly different sensitivity to these;
on the other hand the use of approximate overshooting
in the models of Cohen et al.  (\cite{CCM08}).
The two effects are comparable. 

The difference in the abundances obtained from
the two different MARCS models allow to estimate
the sensitivity of the various abundances to the
model parameters. The difference between the MARCS
and ATLAS overshooting model allow to see the effect 
of the approximate overshooting.
We confirm that [Fe/H] for this star is 0.3 dex
higher using the parameters of Cohen et al.  (\cite{CCM08})
and an ATLAS overshooting model,
relative to that derived using the
parameters of Paper V and a MARCS model (or an 
ATLAS non-overshooting model).
However, 0.17\,dex 
of this difference arises from 
the different choices
in \teff\ and \logg, and 0.13\,dex comes about from
the use of the approximate overshooting.

Having understood these differences we may
conclude that there is excellent agreement between
the two analysis. Note that with our MARCS model
and atmospheric parameters, but the EWs and $gf$ values
of Cohen et al. (\cite{CCM08}),
[Fe/H] for this star is -3.80, which compares very
well with -3.77 given in Paper V.
Note also that when using MARCS models 
(or ATLAS non-overshooting models) our atmospheric
parameters achieve a slightly better iron ionisation
equilibrium (0.03\, dex) than the parameters chosen
by Cohen et al.  (\cite{CCM08}, 0.06\, dex).
However since both these difference are much smaller
than the line-to-line scatter it is impossible to
chose which set of parameters is better by just looking
at the iron ionisation equilibrium.
As noted above most of these differences tend to cancel
out when considering abundance ratios.

Iron is the element for which the largest
number of lines is measured and in this
respect its abundance is more robust.
For other elements the difference between
the values published in Paper V and an
analysis by the 0Z Team may also reflect the
different choice of lines.
For instance for 
BS 16467-062 Cohen et al. (\cite{CCM08}) 
measure 4 \ion{Mg}{i} lines, while
in Paper V we measured 8 lines, but used only
7 to derive the mean Mg abundance.
The Mg lines in BS~16467-062 are all weak; thus, 
the  re-measurement of the Mg abundance  
using line  profile fitting  (see section \ref {mag})
confirms the abundances provided in Paper V.

We discarded \ion{Mg}{i} 416.7271\,nm
because the abundance derived from this line
deviates strongly from those derived from
the other lines. The line is rather weak
(0.75\,pm as measured in our data
or 0.68\, pm as measured by Cohen et al. \cite{CCM08}),
but even for these very weak lines the measurements are 
highly consistent.
Thus the mean Mg abundance from our 7 lines
is, as given in Paper V, 3.97, with a rather
small scatter of 0.09\, dex.
On the other hand the mean Mg abundance from
the four lines measured by  Cohen et al.(\cite{CCM08}),
including  \ion{Mg}{i} 416.7271\,nm, 
and using the atmospheric parameters and model
of Paper V, 
is 4.15 with
a rather large scatter of 0.33\, dex.
The mean Mg abundance for these four lines
from our measurements is 4.12 with a scatter
of 0.38\,dex.
Finally if we take the measurements of  
 Cohen et al.(\cite{CCM08}) and discard
the  \ion{Mg}{i} 416.7271\,nm line
we obtain 3.99 with a scatter of 0.11,
highly consistent with our published value
in Paper V.

The three groups (First Stars, 0Z project, Lai et al.)
have used different
atmospheric parameters for this star,
and the sensitivity of abundances to these
is detailed in all the three papers.
In order to make a stringent comparison
between the results of the three groups
it is advisable to derive abundances from
each set of EWs and $gf$ values for 
a same model atmosphere and with the same
spectrum synthesis code.
We did so in Table \ref{BS_2} in which
we used the MARCS model used in Paper V
to rederive all the abundances.
We compare the atomic species in common, 
excluding Al, for which both ourselves
and Lai et al. have used spectrum synthesis.

Inspection of Table \ref{BS_2} immediately reveals
that, with very few exceptions, the abundances
of the First Stars project rely on a larger number
of lines than those of the other teams. 
This is particularly striking for iron, for which
we use 130 \ion{Fe}{ii} lines compared to 55
of Cohen et al. and 52 of Lai et al.;  a similar
situation is found for Ti, where we use 11 \ion{Ti}{i}
and 23 \ion{Ti}{ii} lines while Cohen et al.
use 2 and 14, respectively and Lai et al. 1 and 9.
This probably reflects the fact that the First Stars
spectra have a larger total wavelength coverage
and a more uniform high S/N ratio across the spectra.
This is due in part to the fact that UVES, as a 
two-arm spectrograph, covers roughly a 30\% larger 
spectral range in a single exposure than HIRES, and
in part to the large amount of telescope time invested in 
the First Stars project.

\begin {table}[t]
\caption {Abundances for BS 16467-062 from
Paper V and  the same model
but
EWs from Cohen et al. (2008), 
 Lai et al. (2008).}
\label {BS_2}
\centering
\begin {tabular}{lrrrrrrrrr}
\hline
\hline
Ion & \multicolumn{3}{c}{EWs } & \multicolumn{3}{c}{EWs } & 
\multicolumn{3}{c}{Paper V}\\
    & \multicolumn{3}{c}{Cohen et al. 2008} &  \multicolumn{3}{c}{Lai et al. 
2008}\\
             &    A &  $\sigma$ & $N$ & A & $\sigma$& $N$ &A & $\sigma$& $N$ \\
\hline
\ion{Mg}{i}  & 3.99$^a$& 0.11 &  3 &  3.98&0.10 & 4 &  3.97 & 0.09 & 7\\  
\ion{Si}{i}  & 4.07    &      &  1 &  4.01&     & 1 &  4.20 &      & 1\\
\ion{Ca}{i}  & 2.71    & 0.12 &  3 &  2.78&0.07 & 4 &  2.94 & 0.19 & 12\\
\ion{Sc}{ii} &--0.54   & 0.03 &  3 & --0.59&0.02& 2 & --0.59 & 0.06 &  4 \\ 
\ion{Ti}{i}  & 1.52    & 0.05 &  2 & 1.57 &     & 1 & 1.65  & 0.17 &  11\\
\ion{Ti}{ii} & 1.42    & 0.11 & 14 & 1.43 &0.12 & 9 & 1.43  & 0.18 &  23\\
\ion{Cr}{ii} & 1.37    & 0.10 &  5 &  1.27&0.07 & 4 & 1.49  & 0.29 &   5 \\
\ion{Mn}{i}  & 0.97    & 0.22 &  5 &  0.82&0.02 & 2 & 1.07  & 0.03 &   3\\
\ion{Fe}{i}  & 3.70    & 0.16 & 55 &  3.61&0.18 &52 & 3.67  & 0.13 & 130 \\ 
\ion{Fe}{ii} & 3.73    & 0.14 &  8 &  3.54&0.16 & 3 & 3.79  & 0.12 &   4\\ 
\ion{Co}{i}  & 1.86    & 0.26 &  3 &  1.66&0.06 & 3 & 1.70  & 0.10 &   4\\
\ion{Ni}{i}  &         &      &  0 &  2.60&0.04 & 2 & 2.56  & 0.03 &   3\\
\hline
\multicolumn{10}{l}{$^a$ line 416.7\,nm has been removed to compute the average}
\end{tabular}
\end{table}

Once the \ion{Mg}{i} line at 416.7\,nm
has been removed from the set of Cohen et al. 
the Mg abundance appears to be in remarkably 
good agreement, in spite of the much larger
number of lines used by the First Stars team.

That the actual choice of lines does make 
a difference is obvious if we look at the Ca 
abundances. There is a difference of 0.23\,dex
in the Ca abundance derived in Paper V and
that of Cohen et al. (2008).
Of the three lines measured by Cohen et al. (2008)
we have only two. The mean Ca abundance for these
two lines is 2.81 with a 0.05\,dex deviation,
thus the discrepancy is reduced to 0.1\,dex, 
totally consistent with the observational errors.
We have measured all four Ca lines used by 
Lai et al. (2008), and the mean 
of these four lines is close to the abundance
given in Table \ref{BS_2}. However,
\ion{Ca}{i} 443.5\,nm appears to be discrepant 
by 0.39\, dex with respect to the mean of the
other three lines, which is  2.86, only
0.08\,dex higher than the value of Lai et al.
and fully consistent with observational
errors.
It is then clear that for the species 
for which a limited number of lines
is available, the actual choice of lines can make a difference.

Another noticeable difference is for Si. All three groups
have determined the Si abundance from a single
Si line, however the other two teams have used
the \ion{Si}{i} 390.6\,nm line, while we
have used the 410.3\,nm line since the other
line is heavily contaminated by CH lines in the spectra
of giant stars. On the other hand, the EWs for
the  390.6\,nm line agree well among the
three investigations (9.18\,pm for us, 9.34\,pm for Cohen et al. 2008
and 9.06\,nm for Lai et al. 2008); thus
the Si abundance derived from this line agrees well 
among the three investigations.

It is reassuring that for iron, for which all three
groups have measured a large number
of lines, the results are fully consistent.

The conclusion of these comparisons is that the 
results of the three teams are consistent,
once the different choice of atmospheric
parameters and models has been factored out.
Some caution must be exercised for the
species which are represented by few lines,
where the actual choice of lines can make
a difference, especially if differential
NLTE effects are present.

\end{document}